\documentclass{aa}  
\usepackage{graphicx}
\usepackage{txfonts}
\usepackage[colorlinks=true,citecolor=blue]{hyperref}
\usepackage{multirow}
\usepackage{amsmath,amstext}
\usepackage[T1]{fontenc}
\usepackage{color}
\usepackage{comment}
\usepackage{caption}
\DeclareCaptionFormat{cont}{#1 (cont.)#2#3\par}
\usepackage[flushleft]{threeparttable}
\usepackage[dvipsnames]{xcolor}  
\usepackage{float}  

\usepackage[switch]{lineno}  

\DeclareRobustCommand{\ion}[2]{%
\relax\ifmmode
\ifx\testbx\f@series
{\mathbf{#1\,\mathsc{#2}}}\else
{\mathrm{#1\,\mathsc{#2}}}\fi
\else\textup{#1\,{\mdseries\textsc{#2}}}%
\fi}

\newcommand{\orcid}[1]{\href{https://orcid.org/#1}{\includegraphics[width=10pt]{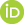}}}

\newcommand{\snia}{SN\,Ia\xspace}
\newcommand{\sneia}{SNe\,Ia\xspace}
\newcommand{\eexsn}{EEx SN\,Ia\xspace}
\newcommand{\eexsne}{EEx SNe\,Ia\xspace}

\newcommand{\snoopy}{SNooPy\xspace}

\newcommand{\tmax}{$t_{0}$\xspace}

\newcommand{\tfd}{$t_{\rm fd}$\xspace}
\newcommand{\tfl}{$t_{\rm fl}$\xspace}
\newcommand{\texp}{$t_{\rm exp}$\xspace}

\newcommand{\g}{\textit{g}\xspace}
\renewcommand{\r}{\textit{r}\xspace}

\newcommand{\nickel}{$^{56}$Ni\xspace}

\newcommand{\chinorm}{$\chi^2_{\rm norm}$\xspace}

\newcommand{\ztfname}{ZTF18abxxssh\xspace}
\newcommand{\smith}{Smith el al. (in prep.)}

\defcitealias{Miller+2020}{M20}
\defcitealias{Deckers+2022}{D22}

\newcommand{\hunits}{km\,s$^{-1}$\,Mpc$^{-1}$\xspace}
\newcommand{\msun}{$M_{\odot}$\xspace}

\begin{document} 

\title{ZTF SN Ia DR2 follow-up: early excess in Type Ia supernova light curves}
\titlerunning{ZTF \sneia Early Excess}
\authorrunning{M\"uller-Bravo et al.}

\author{
    Tomás~E.~M\"uller-Bravo\inst{1, 2}\thanks{\email{t.e.muller-bravo@tcd.ie}}\orcid{0000-0003-3939-7167}
    \and
    Kate~Maguire\inst{1}\orcid{0000-0002-9770-3508}
    \and
    Alaa~Alburai\inst{3, 4}\orcid{0009-0007-2731-5562}
    \and
    Umut~Burgaz\inst{1}\orcid{0000-0003-0126-3999}
    \and
    Georgios~Dimitriadis\inst{5}\orcid{0000-0001-9494-179X}
    \and
    Lluís~Galbany\inst{3, 4}\orcid{0000-0002-1296-6887}
    \and
    Joel~Johansson\inst{6}\orcid{0000-0001-5975-290X}
    \and
    Young-Lo~Kim\inst{7}\orcid{0000-0002-1031-0796}
    \and
    Chang~Liu\inst{8,9,10}\orcid{0000-0002-7866-4531}
    \and
    Adam~A.~Miller\inst{8,9,10}\orcid{0000-0001-9515-478X}
    \and
    Mathew~Smith\inst{11}\orcid{0000-0002-3321-1432}
    \and
    Jesper~Sollerman\inst{6}\orcid{0000-0003-1546-6615}
    \and
    Eric~C.~Bellm\inst{12}\orcid{0000-0001-8018-5348}
    \and
    Joahan~Castaneda~Jaimes\inst{13}\orcid{0000-0002-0987-3372}
    \and
    Mansi~M.~Kasliwal\inst{13}\orcid{0000-0002-5619-4938}
    \and
    Russ~R.~Laher\inst{14}\orcid{0000-0003-2451-5482}
    \and
    Roger~Smith\inst{15}\orcid{0000-0001-7062-9726}
    \and
    Niharika~Sravan\inst{16}
}
\institute{
    School of Physics, Trinity College Dublin, The University of Dublin, Dublin 2, Ireland.
    \and
    Instituto de Ciencias Exactas y Naturales (ICEN), Universidad Arturo Prat, Chile.
    \and
    Institute of Space Sciences (ICE-CSIC), Campus UAB, Carrer de Can Magrans, s/n, E-08193 Barcelona, Spain.
    \and
    Institut d'Estudis Espacials de Catalunya (IEEC), 08860 Castelldefels (Barcelona), Spain.
    \and
    Department of Physics, Lancaster University, Lancaster, LA1 4YB, UK.
    \and
    The Oskar Klein Centre, Department of Astronomy, Stockholm University, AlbaNova SE-10691, Stockholm, Sweden.
    \and
    Department of Astronomy \& Center for Galaxy Evolution Research, Yonsei University, Seoul 03722, Republic of Korea.
    \and
    Department of Physics and Astronomy, Northwestern University, 2145 Sheridan Road, Evanston, IL 60208, USA.
    \and
    Center for Interdisciplinary Exploration and Research in Astrophysics (CIERA), Northwestern University, 1800 Sherman Ave., Evanston, IL 60201, USA.
    \and
    NSF-Simons AI Institute for the Sky (SkAI), 172 E. Chestnut St., Chicago, IL 60611, USA.
    \and
    Department of Physics, Lancaster University, Lancs LA1 4YB, UK.
    \and
    DIRAC Institute, Department of Astronomy, University of Washington, 3910 15th Avenue NE, Seattle, WA 98195, USA.
    \and
    Division of Physics, Mathematics and Astronomy, California 
    Institute of Technology, 1200 E. California Blvd, Pasadena, CA 91125, USA.
    \and
    IPAC, California Institute of Technology, 1200 E. California Blvd, Pasadena, CA 91125, USA.
    \and
    Caltech Optical Observatories, California Institute of Technology, Pasadena, CA  91125, USA
    \and
    Department of Physics, Drexel University, Philadelphia, PA 19104, USA.
}

\date{Received \today; accepted ---}

\abstract
{
There is a general consensus that Type Ia supernovae (\sneia) are the thermonuclear explosions of a C/O white dwarf (WD) in a binary system. However, the nature of their progenitors and explosion mechanism still remains elusive.
Their early light curves can offer strong constraints on these as they trace the outermost layers and provide evidence of their initial conditions. 
In this work, we study the Zwicky Transient Facility data release 2 sample of \sneia in search for objects with early flux excess (\eexsne). 
Two main approaches are used: i) finding deviations between early-time observations and models without early excess, and ii) agreement between observations and models with early excess. 
For the former approach, different methods, such as power-law and SALT2 fits, were used to identify \eexsn candidates. For the second approach, we used double-detonation (DD) and companion-interaction (CI) models, finding 145 and 199 candidates, respectively. In general, there is a disagreement between the methods in the exact objects identified as \eexsne and also their number.
Combining multiple methods, we find 17 strong \eexsn candidates, of which six are of over-luminous subtype (five 91T-like and one 03fg-like). The most prominent excesses reach ${\sim}10-17\%$ of peak flux (up to ${\sim-}17$\,mag). We found that the best \eexsn candidates have higher stretch on average compared to the rest of the \sneia. They also seem to prefer lower-mass and bluer host galaxies, potentially linked to younger stellar populations. Additionally, we estimate the fraction of \eexsne to be $\lesssim25\%$ of all \sneia, in agreement with previous works, with over-luminous subtypes having a higher relative fraction compared to normal \sneia.
When studying the parameter distribution of the DD models for the \eexsn candidates, we see a preference for intermediate WD core mass ($1$\,\msun), a lower fraction of shell burning ($20\%$) and heavier He-burning products (\nickel). For the CI models, we see no trends with any of the parameters. We are unable to draw strong conclusions on whether one model is preferred over the other as a viable description of \snia explosions
Finally, we highlight the challenge of identifying \eexsne in \lq real time', making the triggering of early-time spectroscopic observations for these objects a difficult task.
}

\keywords{supernovae: general}
\maketitle

\section{Introduction}

Type Ia supernovae (SNe Ia) are transient astrophysical events believed to originate from the thermonuclear explosion of a C/O white dwarf (WD) in a binary system. In the canonical scenario, the WD accretes mass from its companion star until it approaches the Chandrasekhar mass limit ($\mathord{\sim}1.4$\,\msun), triggering runaway nuclear burning and ending in a very energetic explosion \citep[see reviews by][]{Maoz+2014, Livio_Mazzali+2018}. To date, two primary progenitor channels have been proposed: (i) the single-degenerate model, where accretion proceeds from a non-degenerate star \citep{Whelan_Iben1973, Nomoto+1984}, and (ii) the double-degenerate model, in which two WDs merge due to orbital decay via gravitational-wave emission \citep{Tutukov_Yungelson1981, Iben_Tutukov1984, Webbink1984, Wang_Han2012}. Despite decades of study, the dominant progenitor pathway remains uncertain.

Observations of early-time light curves provide crucial diagnostics of the explosion mechanism and progenitor system. 
First, the interval between the explosion and shock breakout, provided the explosion epoch can be constrained, yields a measure of the pre-explosion stellar radius. This \lq dark phase' persists until photons from radioactive heating diffuse outward, with a duration that depends on the depth of radioactive material and can extend up to a few days; for instance, a $\mathord{\sim}1$\,day dark phase was inferred for SN 2011fe \citep{Mazzali+2014}. On the other hand, the shock breakout has been observed in some core-collapse SNe \citep[e.g.,][]{Tominaga+2011, Garnavich+2016}, although for \sneia the typical shock breakout flash is expected to be much weaker and shorter (milliseconds to seconds), given that they are thought to come from compact stars (e.g., WD), making it very difficult to detect. Nonetheless, non-detections of the associated thermal emission can place stringent upper limits on the progenitor radius and therefore the nature of it
(e.g., \citealt{Bloom+2012, Brown+2012b}; although see \citealt{Brown+2012a} for constrains from early UV observations as well). 
Second, the rise of the light curve, powered by the decay of \nickel \citep{Colgate_McKee1969, Arnett1982}, reflects the radial distribution of radioactive material in the ejecta, thereby probing ignition conditions, burning processes, and ejecta mixing. The light-curve rise of \sneia has been commonly parametrised by the so-called \lq fireball' model, which describes the rise-time luminosity as a power law with an index of $2$ ($L_{\rm opt} \propto t^{2}$; \citealt{Nugent+2011}). However, \citet{Piro_Nakar2013} showed that this model is generally unrealistic for describing all \sneia, as it neglects the time dependence of the photospheric velocity and assumes a uniform radial distribution of \nickel. Multi-dimensional explosion models have similarly shown that deviations from $t^{2}$ are expected \citep{Dessart+2014, Magee+2020}. 
Third, at early times (hours to days post-explosion), additional excess emission may arise from processes such as ejecta interaction with a companion star, through Roche-lobe overflow, or circumstellar material, typically manifesting as a blue/UV \lq bump' that is particularly pronounced within the first few days \citep{Kasen2010, Hayden+2010b, Goobar+2015, Hosseinzadeh+2017, Jiang+2017, Kumar+2025}. 

Spectroscopic observations of infant \sneia provide information about the outer ejecta and its chemical structure, constraining the nature of the accreting surface layer. For instance, the explosive He burning in the double-detonation scenario may leave traces of He and produce some amount of iron-group elements \citep[e.g., $^{44}$Ti, $^{48}$Cr, $^{52}$Fe, $^{56}$Ni;][]{Fink+2010, Jiang+2017, Liu2023, Padilla-Gonzalez+2024}. On the other hand, delayed-detonation models predict C, O, and possibly Si detectable up to a couple of days after the explosion \citep{Mazzali2001, Blondin+2013, Hoeflich+2017}. Models involving interaction with a non-degenerate companion predict an early UV/blue continuum excess; however, this and other observational signatures are difficult to distinguish from those produced by alternative scenarios. \citep{Kasen2010, Marion2016}. Therefore, surveys such as KMTNet SN Program \citep{Moon+2016}, DLT40 \citep[][]{Sand+2018}, the Young Supernova Experiment \citep[YSE;][]{Jones+2021}, and the Precision Observations of Infant Supernova Explosion \citep[POISE;][]{Burns+2021}, which focused on the early discovery and follow-up of transients, play a crucial role in constraining progenitor scenarios of \sneia. However, the number of \sneia with early-time spectroscopic observations still remains relatively limited \citep[for some recent works, see e.g.][]{Galbany+2025, Hoogendam+2025, Iskandar+2025, Ni+2025}.

High-cadence, wide-field photometric surveys are essential for capturing infant transients. Facilities such as the Asteroid Terrestrial-impact Last Alert System \citep[ATLAS;][]{Tonry+2018, Smith+2020} and the Zwicky Transient Facility \citep[ZTF;][]{Bellm+2019, Graham+2019, Masci+2019, Dekany+2020} constantly monitor the night sky with a cadence of a few days, discovering thousands of transients every year. Although ATLAS has a larger sky coverage, ZTF goes much deeper and is thus better suited for early-time observations. 
In this work, we focus on identifying \sneia with early flux excess \citep[\eexsne;][]{Jiang+2018} from the ZTF data release (DR) 2 sample and compare different methods for their identification. In addition, we study some of the leading theoretical models that predict early flux excess in \sneia to place some constrains on their progenitor scenario.
The structure of this paper is as follows: in Sect.~\ref{sec:data} we briefly describe the ZTF DR2 sample and the selection cuts applied to it; the methods used for identifying early flux excess are described in Sect.~\ref{sec:methods}; the results of the \eexsn identification are presented in Sect.~\ref{sec:results}; in Sect.~\ref{sec:discussion}, we discuss the results and compare the different methods used for identifying \eexsne, and also analyse the theoretical models with early flux excess used in this work; finally, the conclusions are presented in Sect.~\ref{sec:conclusions}. 

\section{Data}
\label{sec:data}

In this section, we describe the \snia sample used and the processing of the light curves needed for the analysis.

\subsection{Sample selection}
\label{subsec:sample}

In this work, we use the ZTF DR2 sample which consists of 3,628 spectroscopically classified \sneia \citep[][]{Rigault+2025a}. These SNe were observed in \textit{gri}-band, but only a relatively small fraction of these have good \textit{i}-band coverage (\smith). The light curves were parsed using the \texttt{ztfidr}\footnote{\url{https://github.com/MickaelRigault/ztfidr}} package. A flux-dependent error floor of $2.5\%$, $3.5\%$ and $6\%$ in \textit{gri}-band, respectively, is added in quadrature to the flux uncertainties \citep[][\smith]{Rigault+2025a, Amenouche+2025}. We also apply the recommended cuts to remove epochs with inaccurate photometry caused by different issues, such as bad weather and incorrect background subtraction\footnote{Flags 1, 2, 4, 8, 16 and 1065.}.

We retrieve SALT2 \citep{Guy+2005, Guy+2007} light-curve parameters for the ZTF DR2 sample also using \texttt{ztfidr}. In particular, the time of optical peak (\tmax) is used for obtaining different phase ranges used in this work (see Sect.~\ref{sec:methods}), but other light-curve parameters (Sect.~\ref{subsec:parameters_distributions}) and the light-curve model (Sect.~\ref{subsec:salt2}) are used as well. Initially, we apply the suggested light-curve coverage cut (\texttt{lccoverage\_flag=1}), reducing the initial sample from 3,628 to 2,960 \sneia (see Table~\ref{tab:cuts}). The fit-quality (\texttt{fitquality\_flag=1}) cut is later used in the analysis to study the light-curve parameter distributions (Sect.~\ref{subsec:parameters_distributions}).

\begin{table}[h]
\caption{Sample selection.}
\centering
\begin{threeparttable}

\begin{tabular}{llc}
\hline
Cut & Remaining & Removed \\
\hline
Before cuts & 3628 & - \\
\texttt{lccoverage\_flag=1} & 2,960 & 668 \\
\tfd calculated & 2,813 & 147 \\
Two \textit{g}-band epochs in [\tfd, \tfd $+ 5$]\,d & \textbf{1,792} & 1,021 \\
\hline
\end{tabular}

\end{threeparttable}
\label{tab:cuts}

\end{table}

\subsection{Light-curve processing}
\label{subsec:lc_processing}

Ideally, rest-frame light curves are required for identifying flux excess at early epochs, for which $K$-correction, and therefore knowledge of the spectral energy distribution (SED), is needed \citep{Oke_Sandage1968, Kim+1996, Hsiao+2007}. \cite{Bulla+2020} argues that $K$-corrections at early epochs are small as suggested by the fits to the light curves of \sneia using the \snoopy light-curve fitter \citep[][]{Burns+2011, Burns+2014}. However, one should note that light-curve fitters are trained on limited amount of data or rely on extrapolations at such epochs, given the difficulty of having early-time follow-up observations. Additionally, the SED for different \snia subtypes can vary significantly \citep[e.g.][]{Dimitriadis+2025}. We therefore prefer not to include $K$-corrections due to the uncertainty of these. However, note that due to the relatively low redshift of the ZTF DR2 sample ($z\lesssim0.1$), $K$-corrections are not expected to affect our analysis significantly \citep{Liu_Miller2006, Liu+2026}. As redshift increases, SN-frame UV flux shifts towards observer-frame optical bands, in principle making the detection of early excess much easier, although the coverage at early epochs also becomes challenging due to the magnitude limit of the survey ($\sim21$\,mag). 

We apply Milky-Way dust extinction correction to the SN light curves using the dust maps from \cite{Schlafly+2011} and the dust extinction law from \cite{Fitzpatrick+1999}. The light curves are also corrected for time dilation due to the expansion of the Universe using the heliocentric redshifts associated to the \sneia.

A step required for our analysis is to estimate the epoch\footnote{The term `epoch' is used to denote a single night of observations since some nights include multiple observations.} at which the SN first becomes detected by ZTF. We refer to this epoch as the time of first detection (\tfd). For this, we calculated nightly weighted-average fluxes for each available band, with lower phase bound of $25$ rest-frame days before optical peak brightness (\tmax). This lower phase limit avoids sporadic false detections at earlier epochs, which are not expected for \sneia, and also covers their observed rise time range \citep{Hayden+2010a, Firth+2015, Miller+2020}. We use three approaches, in following order, where the subsequent approach is only used if the previous one does not yield a \tfd value:

\begin{itemize}
    \item Selecting the first epoch with a $3\sigma$ detection (above zero flux), with a subsequent $3\sigma$ epoch and two prior non-detections epochs (<$3\sigma$).
    \item Going "backwards" from \tmax into earlier epochs, selecting the last-detection epoch, with two consecutive non-detections epochs.
    \item Selecting the first detection epoch, with a subsequent detection.
\end{itemize}

These approaches go in decreasing order of accuracy for the estimation of \tfd, but none of these is immune to flagging the incorrect epoch. The most robust approach, the first one, yields values for most objects, while the three combined produce values for 2,813 of the 2,960 \sneia.

\begin{figure}[h!]
    \includegraphics[width=\columnwidth]{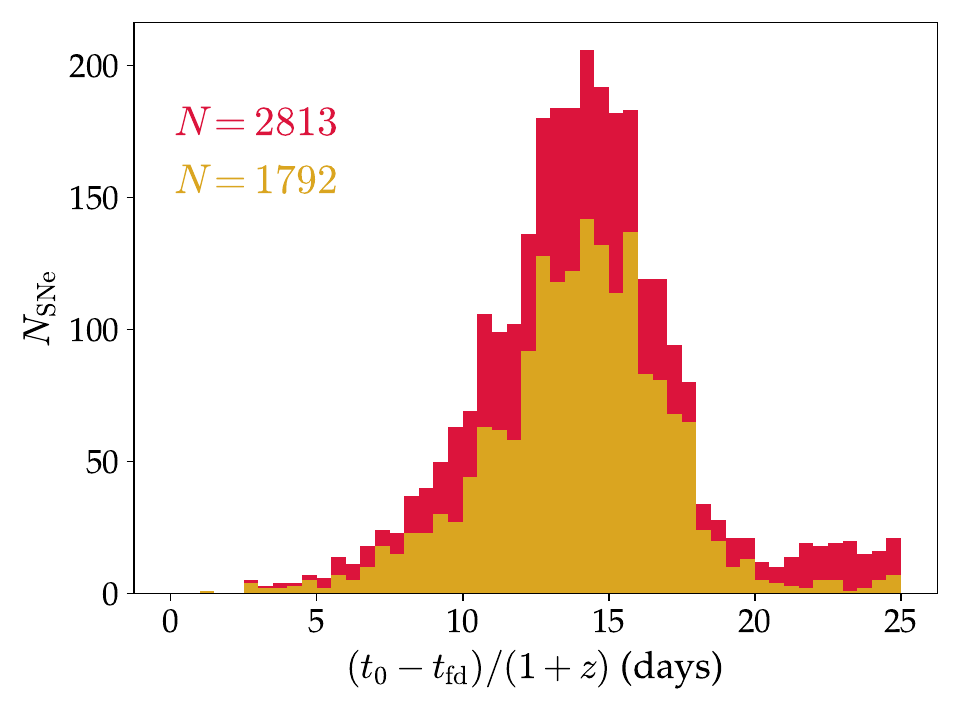}
    \caption{Pseudo-rise time, i.e. rest-frame time from \tfd to \tmax, of 2,813 out of 2,960 \sneia from the ZTF DR2 sample (red). 1,792 objects have at least two \textit{g}-band epochs in the range [\tfd, \tfd $+ 5$]\,d (golden). Those with relatively short rise times ($\lesssim 12$\,d) are mostly fast-evolving objects (e.g., 91bg-like or Iax SNe) or lack early-enough detections due to gaps in the observations or apparent faintness. See Sect.~\ref{subsec:lc_processing} for more details.}
    \label{fig:rise_from_tfd}
\end{figure}

We calculate the rest-frame pseudo-rise time of the \sneia in the sample (Fig.~\ref{fig:rise_from_tfd}), defined as $(t_0 - t_{\rm fd}) / (1 + z)$. Objects with less than two \textit{g}-band epochs in the range [\tfd, \tfd $+ 5$]\,d are removed from the sample, as this is a minimum requirement for our analysis (Sect.~\ref{sec:methods}). This last cut reduces the sample to 1,792 \sneia. In Table~\ref{tab:cuts}, we summarise the cuts applied in this work.

\section{Methods}
\label{sec:methods}

In this section, we present the methods and criteria used for identifying \eexsne.

\subsection{Excess detection}
\label{subsec:excess}

The interaction of the \snia ejecta with a companion star is expected to last from hours to a few days after the SN explosion \citep[e.g.][]{Kasen2010, Magee+2021}. Similarly, double-detonation models predict early excess from radioactive decay lasting about the same amount of time \citep{Magee+2022}. However, since the rise time of \sneia can vary significantly between individual objects \citep[e.g.][]{Hayden+2010a, Firth+2015, Miller+2020}, and given the diversity of the early excess duration, the estimation of the time frame at which the flux excess is expected becomes challenging. Therefore, we explore a range of times, using the time of first detection (see Sect.~\ref{subsec:lc_processing}) as a lower bound and different upper bounds for detecting early flux excess. We will refer to this time range as the \textit{excess window}.

To identify early excesses in the light curves of \sneia, we use two main approaches: i) fitting light-curve models without flux excesses and checking for deviations between these and the observations in the excess window, and ii) fitting models that include early flux excesses and checking whether these are favoured over models without excesses. 

For the first approach, we use four different methods to search for early excess in \snia light curves: i) comparison to SALT2 model light curves (Sect.~\ref{subsec:salt2}), ii) comparison to a simple power-law model (Sect.~\ref{subsec:powerlaw}), iii) comparison to a modified power-law model (Sect.~\ref{subsec:powerlaw}), and iv) comparison to radiative transfer models of \cite{Magee+2020} varying the \nickel\ distribution in the ejecta (Sect.~\ref{subsubsec:ni_dist}). We emphasise that these methods are optimised to detect early excess and can differ from similar approaches used for estimating \snia rise properties. In addition, we make use of the normalised light-curve residual in units of errors, \texttt{pull = (data - model) / error}, for identifying whether data points are consistent or not with a given model. For the identification of a flux excess, we require to have at least two different nights in the excess window with data deviating from a model without flux excess, with a significance of \texttt{pull} $\geq 5$. As the flux excess is primarily expected at bluer wavelengths, we focus on the $g$-band light curves, although we also investigate possible excesses in the $r$ band for a small subset of objects.

For the second approach, described in Sect.~\ref{subsubsec:dd_ci}, we simply fit models with and without early flux excess to the \sneia in our sample and determine if there is a statistical preference for the former or the latter.

\subsection{SALT2 model}
\label{subsec:salt2}

We gather the SALT2 (version \textsc{T21}; \citealt{Taylor2021}) light-curve fits from ZTF DR2 using \texttt{ztfidr}. The phase range used for the fits include $\phi \in [-10, +40]$\,d, with respect to \tmax. \cite{Rigault+2025b} found that the SALT2 model has excellent agreement with the data, on average, in the phase range $\phi < -10$\,d, despite being trained on limited spectro-photometric data at these phases (the model extrapolates epochs earlier than $\mathord{\sim}-20$\,d). However, they also found that the model over-predicts, on average, the luminosity of \sneia around $\phi\sim-15$\,d, which can partially coincide with the excess window. This is most likely a bias caused by the limited number of \sneia used for the training of SALT2 at early epochs \citep{Rigault+2025a}. Note that the inclusion of \eexsne in the training would produce this effect, but we believe this to be unlikely given the lack of training data at these phases.

\begin{figure}[h!]
    \includegraphics[width=\columnwidth]{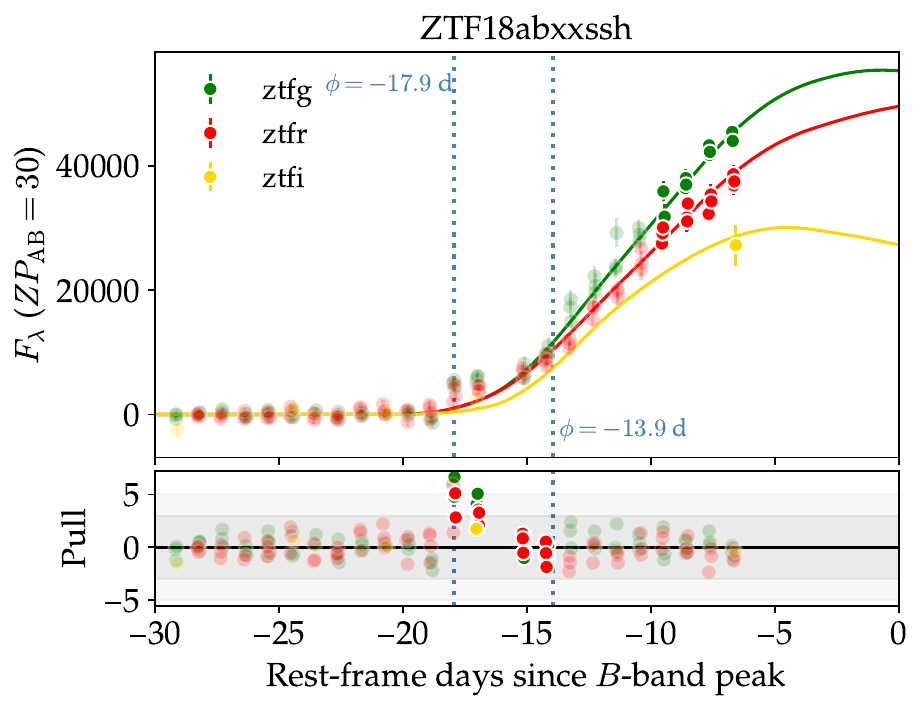}
    \caption{ZTF \textit{gri} rest-frame light curves of \ztfname. \textit{Top}: opaque circles are epochs included in the SALT2 fit, while semi-transparent circles are epochs not included (see Sect.~\ref{subsec:salt2}). Solid lines represent the best-fitting SALT2 model. The vertical dotted blue lines mark the bounds of the \lq excess window\rq ($\phi \in [t_{\rm fd},\, t_{\rm fd} + 5]$\,d in this case). \textit{Bottom}: semi-transparent circles in $\phi \in [t_{\rm fd}\,+\,5, -10]$\,d (i.e. to the right of ther upper bound) are epochs used for checking whether the SALT2 model is under-predicting the light curve flux, while opaque circles are epochs used for identifying flux excess. The $3\sigma$ and $5\sigma$ regions are marked with different shades of grey. This SN is identified as an \eexsn with this method.}
    \label{fig:salt2_example}
\end{figure}

As the fit range (i.e. $[-10, +40]$\,d) does not include the excess window, early bumps are not expected to bias the light-curve fits. In addition, we require the observations between the upper bound of the excess window and the beginning of the fit range (i.e. up to $-10$\,d) to agree with the SALT2 model within $3\sigma$. This helps exclude SNe with over- or under-predicted flux, avoiding false excess detection in the latter case. An example of the SALT2 model comparison to the pre-maximum light curve of a SN Ia (\ztfname) displaying an early flux excess is shown in Fig.~\ref{fig:salt2_example}.

\subsection{Power-law models}
\label{subsec:powerlaw}

In this section, we follow a similar approach as in \citet[][hereafter \citetalias{Miller+2020}]{Miller+2020} to fit a multi-band power-law to the SN light curves:

\begin{equation}
    \hspace{2.5cm} f_b(t) = C + H[t_{\rm fl}]\,A_b\,(t - t_{\rm fl})^{\alpha_b},
    \label{eq:power_law}
\end{equation}

where $C$ is the baseline of the light curves, \tfl is the time of first light (assumed to be the same for all bands, as in \citetalias{Miller+2020}; not to confuse with the time of first detection), $H[t_{\rm fl}]$ is the Heaviside function equal to $0$ for $t<$\tfl and $1$ otherwise, $A_b$ is the flux scale in filter $b$, and $\alpha_b$ is the power-law index in filter $b$. In our case, $C$ is fixed at zero flux given that the baseline was already subtracted from the SN light curves, and the uncertainty scale factor used by \citetalias{Miller+2020} is not added as the error floor should account for this. Another difference is that we exclude epochs that fall in the excess window to avoid fitting through the excess. Only $gr$-band light curves are fitted with a power law as the $i$ band is sparsely covered in general and excess is not expected at the wavelength range covered by this band.

Given the limited data available in the fit range, some further changes are made: data up to $50\%$ of peak $g$-band flux is used (\citetalias{Miller+2020} uses data up to $40\%$ of peak flux), and the priors on the power-law index and \tfl are changed with respect to those in \citetalias{Miller+2020} to better constrain the fits. A summary of the priors is presented in Table~\ref{tab:priors}. Note that we also use the change of variable $A' = A 10^{\alpha}$. 

\begin{table}[ht!]
\caption{Power-law model parameters and their priors.}
\centering
\begin{threeparttable}

\begin{tabular}{llc}
\hline
Parameter & Description & Prior \\
\hline
$t_{\rm fl}$ & Time of first light & $\mathcal{U}(-100, t_{\rm{fd}})$ \\
$A'^b$ & Scale per filter & $(A'^{b})^{-1}10^{-\alpha^b}$ \\
$\alpha^b$ & Power-law index & $\mathcal{U}(1, 10^8)$ \\
$\beta^b$ & 2nd power-law index & $\mathcal{U}(-10^8, 10^8)$ \\
\hline
\end{tabular}

\begin{tablenotes}
 \item \textbf{Notes.} The priors for $t_{\rm fl}$ are rest-frame days with respect to \tmax. The second power-law index, $\beta_b$, is used only for the modified power law.\\ 
\end{tablenotes}

\end{threeparttable}
\label{tab:priors}

\end{table}

\begin{figure}[h!]
    \includegraphics[width=\columnwidth]{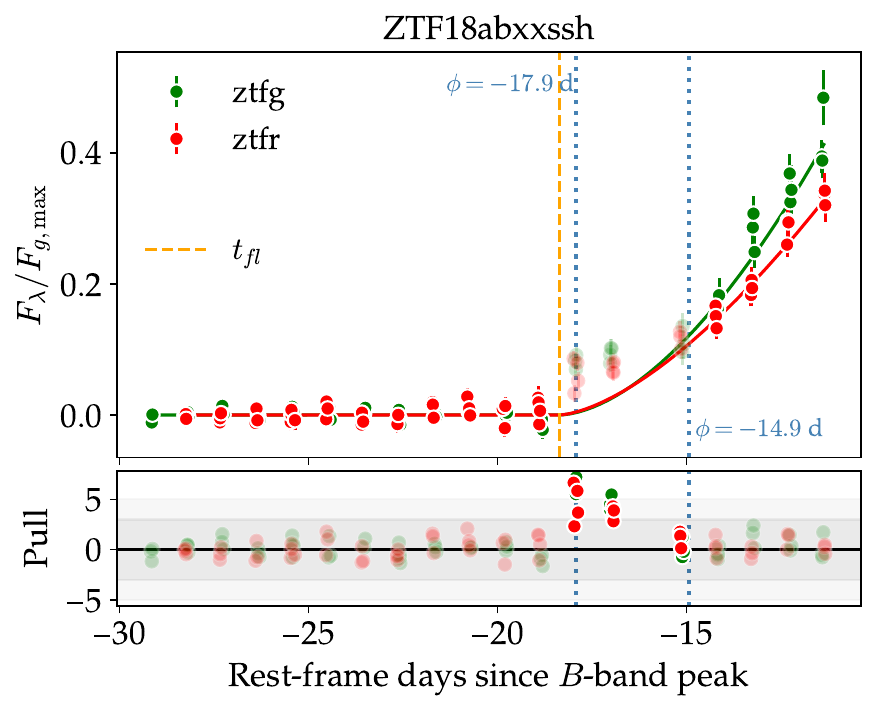}
    \caption{ZTF \textit{gr} rest-frame light curves of \ztfname. \textit{Top}: opaque circles are epochs included in the power-law fit (see Sect.~\ref{subsec:powerlaw}), while semi-transparent circles are epochs not included as they fall in the excess window. The vertical lines mark the time of first light (\tfl; dashed orange line) and the excess window boundaries ($\phi \in [t_{\rm fd},\, t_{\rm fd} + 3]$\,d in this case; dotted blue lines). \textit{Bottom}: opaque circles fall in the excess window and are used for identifying flux excess. The $3\sigma$ and $5\sigma$ regions are marked with different shades of grey. This SN is identified as an \eexsn with this method.}
    \label{fig:powerlaw_example}
\end{figure}

The Markov Chain Monte Carlo (MCMC) technique is used to fit the model to the SN light curves, with 100 walkers and 2,000 steps each. An example fit of this method to a SN Ia (ZTF18abxxssh) showing an early flux excess is shown in Fig.~\ref{fig:powerlaw_example}.

In addition to the power-law model, we also include a `modified' power-law model as in \cite{Vallely2021}:

\begin{equation}
    \hspace{2.5cm} f_b(t) = C + H[t_{\rm fl}]\,A_b\,(t - t_{\rm fl})^{\alpha_b(1 + \beta_b(t - t_{\rm fl}))},
    \label{eq:power_law}
\end{equation}

where the $\beta_b$ index allows the model to follow the light curve at later times up to peak flux. This presents an advantage over the former model as more SNe can be included. The fitting of the modified power-law model follows the same prescription used for the simple power law.

\subsection{TURTLS models}
\label{subsec:turtls}

In this section, we use the models produced with the TURTLS radiative transfer code \citep{Magee+2018} to identify early excess in the \snia light curves. Note that the models used in Sect.~\ref{subsubsec:ni_dist} do not show early flux excess, while some of those used in Sect.~\ref{subsubsec:dd_ci} do.

\subsubsection{\nickel-distribution models}
\label{subsubsec:ni_dist}

Here, we follow a similar approach as in \citet[][hereafter \citetalias{Deckers+2022}]{Deckers+2022} and use the models presented in \cite{Magee+2020}\footnote{This includes the 45 models produced for \citetalias{Deckers+2022}.} to look for early excess. These 300 models explore varying distributions of \nickel in the SN ejecta as a function of kinetic energy ($0.50-2.18\times10^{51}$\,erg), with different density profiles (double power law [DPL] and exponential [EXP]), and other parameters. We will refer to these as the \nickel-distribution models or \nickel-distribution method.

As in \citetalias{Deckers+2022}, only two nuisance parameters are fit with the models: distance modulus ($\mu$) and explosion epoch (\texp). As an initial guess for $\mu$, we use the SN redshift assuming a flat $\Lambda$ Cold Dark Matter (CDM) cosmology with $H_0=70$\,\hunits and $\Omega_m=0.3$. For \texp, we use half a day before \tfd as the time of explosion happens before first detection. We do not expect these initial guess values to impact the fits in any significant way. Similar bounds for the distance and explosion epoch parameters as in \citetalias{Deckers+2022} are used as well: for the distance, $\mu\pm0.3$\,mag\footnote{Note that a difference of $1$\,\hunits in $H_0$ correspondences to $\mathord{\sim}0.03$\,mag in distance modulus.}, and for \texp, $[t_{\rm fd} - 30,\, t_{\rm fd}]$ rest-frame days. As \texp$\geq$\tfd is unphysical, we changed the upper boundary compared to \citetalias{Deckers+2022}.

\begin{figure}[ht!]
    \includegraphics[width=\columnwidth]{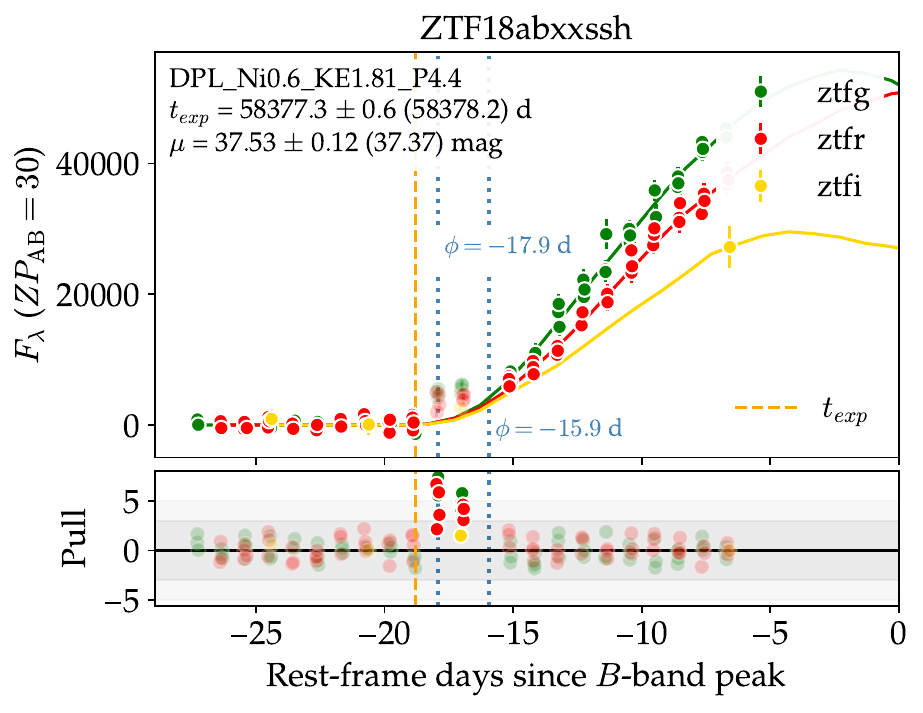}
    \caption{ZTF \textit{gri} rest-frame light curves of \ztfname. \textit{Top}: best-fit \nickel-distribution model from Sect.~\ref{subsubsec:ni_dist} for this SN. Best fit values are also shown in the plot with the initial values in parentheses. Semi-transparent circles are epochs not included in the fit as they fall in the excess window. \textit{Bottom}: best-fit light-curve residual (i.e. pull). Opaque circles are epochs in the excess window used to identify flux excess. The vertical lines mark the epoch of explosion, \texp (dashed orange line) and the excess window boundaries ($\phi \in [t_{\rm fd},\, t_{\rm fd} + 2]$\,d in this case; dotted blue lines). The $3\sigma$ and $5\sigma$ regions are marked with different shades of grey. This SN is identified as an \eexsn with this method.}
    \label{fig:turtls_example}
\end{figure}

A major difference with the method in \citetalias{Deckers+2022} is that we tested excluding data falling in the excess window. As \cite{Magee+2020} showed, early data is essential to avoid biasing the fits of the \nickel-distribution models. However, it is also where the flux excess is expected to occur. The inclusion of baseline data from \tfd$- 10$\,d onward should anchor the model, whilst varying the exclusion time range (i.e. where data is removed) allows us to identify the optimal range for fitting the models and detecting flux excess. In summary, data with $\phi \in [t_{\rm fd}-10, t_0]$\,d is considered in the fit and the flux excess detection is done as with the previous methods above. In this case, only the best-fit model is considered. An example is shown in Fig.~\ref{fig:turtls_example}.

\subsubsection{Double-Detonation and Companion-Interaction models}
\label{subsubsec:dd_ci}

In this section, we describe the second approach mentioned in Sect.~\ref{subsec:excess} (i.e. using models with early excess to fit the light curves), and consider a separate analysis for this. We use 350 double-detonation (DD) models from \cite{Magee+2021} and 96 companion-interaction (CI) models from \cite{Magee+2022}, both produced with the TURTLS radiative transfer code.
The DD models simulate SN ejecta with varying 
WD core masses ($0.9-1.2\,\text{M}_{\odot}$), 
He shell masses ($0.01-0.1\,\text{M}_{\odot}$), 
burning shell fraction ($20-80\%$), 
and dominant product from He burning ($^{32}\text{S} \rightarrow ^{56}$Ni).
The CI models simulate SN ejecta with varying 
kinetic energy ($0.65-1.68\times10^{51}$\,erg), 
\nickel-distribution scaling parameter ($s$; $3, 9.7$), 
companion separation ($3-200\times10^{11}$\,cm), 
and viewing angle ($0-135^{\circ}$).
A summary of the number of TURTLS models included in this work is shown in Table~\ref{tab:turtls}.

\begin{table}[h]
\caption{TURTLS models \citep{Magee+2018} used in this work.}
\centering
\begin{threeparttable}

\begin{tabular}{lcc}
\hline
Model Type & \# of models & Reference \\
\hline
\nickel distribution & 255 & \cite{Magee+2020} \\
 & 45 & \citetalias{Deckers+2022} \\
Double detonation & 350 & \cite{Magee+2021} \\
Companion interaction & 96 & \cite{Magee+2022} \\
\hline
All & 746 & \\
\hline
\end{tabular}

\begin{tablenotes}
 \item \textbf{Notes.} From the double-detonation models, 157 show early excess and 193 do not, while from the companion-interaction models, 48 show early excess and 48 do not.
\end{tablenotes}

\end{threeparttable}
\label{tab:turtls}

\end{table}

\begin{figure}[h!]
    \includegraphics[width=\columnwidth]{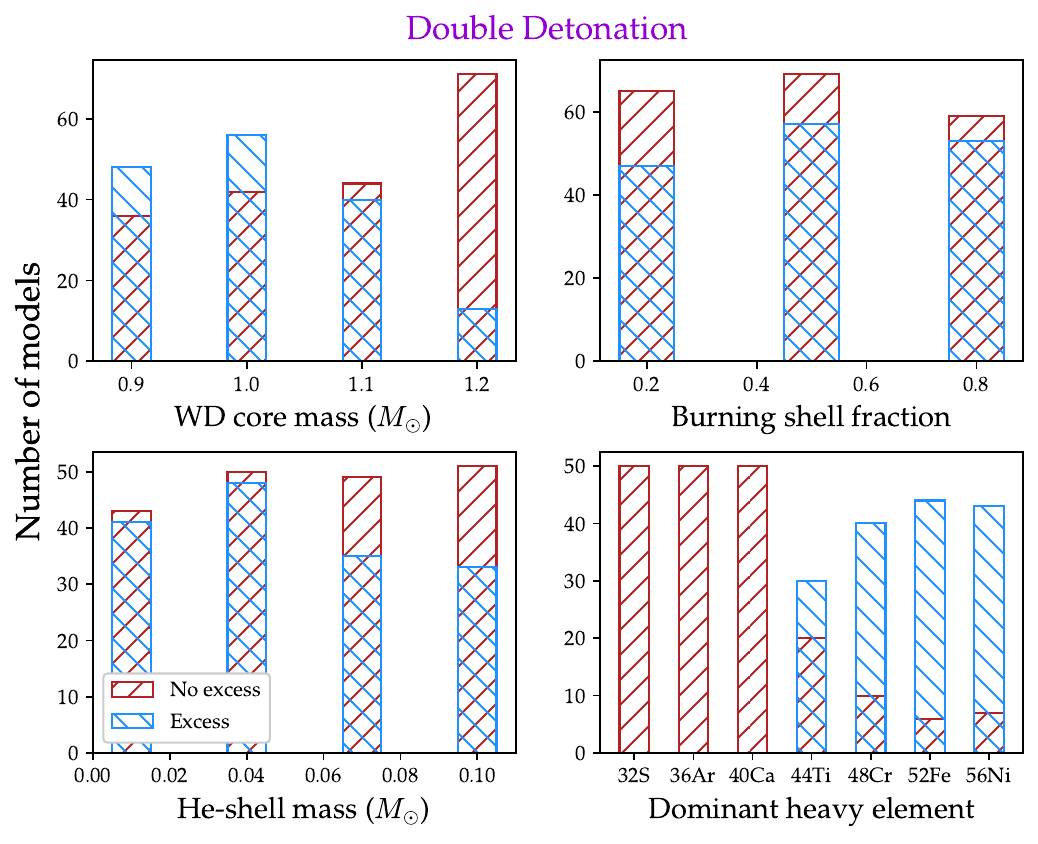}
    \caption{Parameter distributions for the \textcolor{Plum}{double-detonation} models from \cite{Magee+2021}. In blue and red are the models with and without early flux excess, respectively.}
\label{fig:parameter_distributions_all_DoubleDetonation}
\end{figure}

In contrast to the \nickel-distribution models, some of the DD and CI models do show clear early flux excess in their light curves. These were identified by visual inspection, where any model showing either a clear `shoulder' or `bump'\footnote{This is purely a qualitative description of the early excess and we do not attempt to quantify the shape or strength of it as part of this work. We consider any deviation from a smooth rising light curve to be an excess.} during the early rise were considered as having early excess. From the CI models, those with $s=3$ (i.e. large fraction of \nickel in the outer ejecta) do not show signs of early excess, while all of those with $s=9.7$ do. Half (48) of the models show early excess while the other half do not. In the case of the DD models, those with lighter intermediate-mass elements as He-burning product, such as $^{32}$S, $^{36}$Ar and $^{40}$Ca, do not show early excess, but also some of the other models with heavier elements as well (particularly $^{44}$Ti; see Fig.~\ref{fig:parameter_distributions_all_DoubleDetonation}). 157 ($45\%$) of the models show early excess and 193 ($55\%$) do not. 

\begin{figure}[h!]
    \includegraphics[width=\columnwidth]{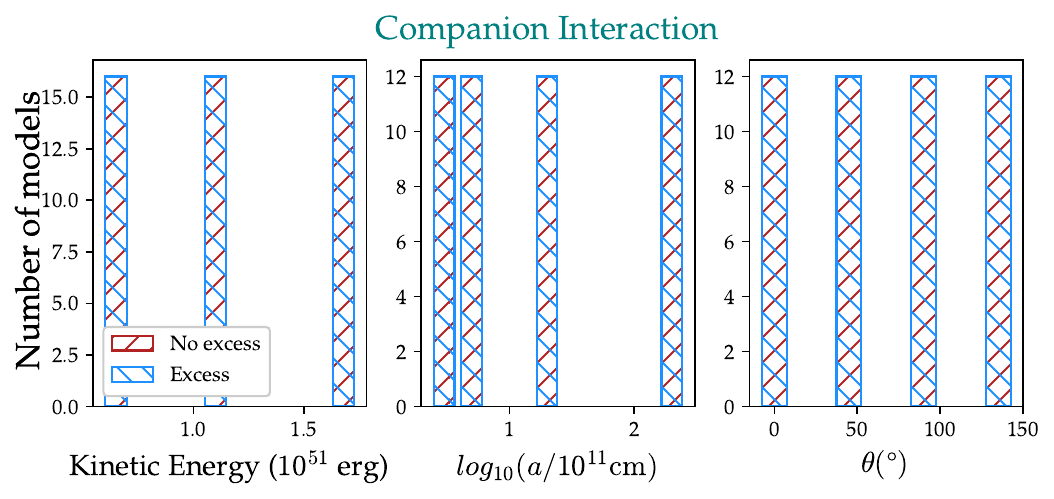}
    \caption{Parameter distributions for the \textcolor{teal}{companion-interaction} models from \cite{Magee+2022}. In blue and red are the models with ($s=9.7$) and without ($s=3$) early flux excess, respectively. Note that the parameter distributions are the same between those models with and without early excess.}
    \label{fig:parameter_distributions_all_CompanionInteraction}
\end{figure}

The parameter distribution of the DD and CI models, with and without early excess, are shown in Figs.~\ref{fig:parameter_distributions_all_DoubleDetonation} and \ref{fig:parameter_distributions_all_CompanionInteraction}. It can be seen that for the DD models, an excess is seen mainly for WD core masses below $1.2$\,\msun, and for dominant iron-group elements at least as heavy as $^{44}$Ti. For the CI models, the \nickel-distribution scaling parameter, $s$, is the sole source of the flux excess: models with $s=3$ do not show early excess, while models with $s=9.7$ do.

\begin{figure}[h!]
    \includegraphics[width=\columnwidth]{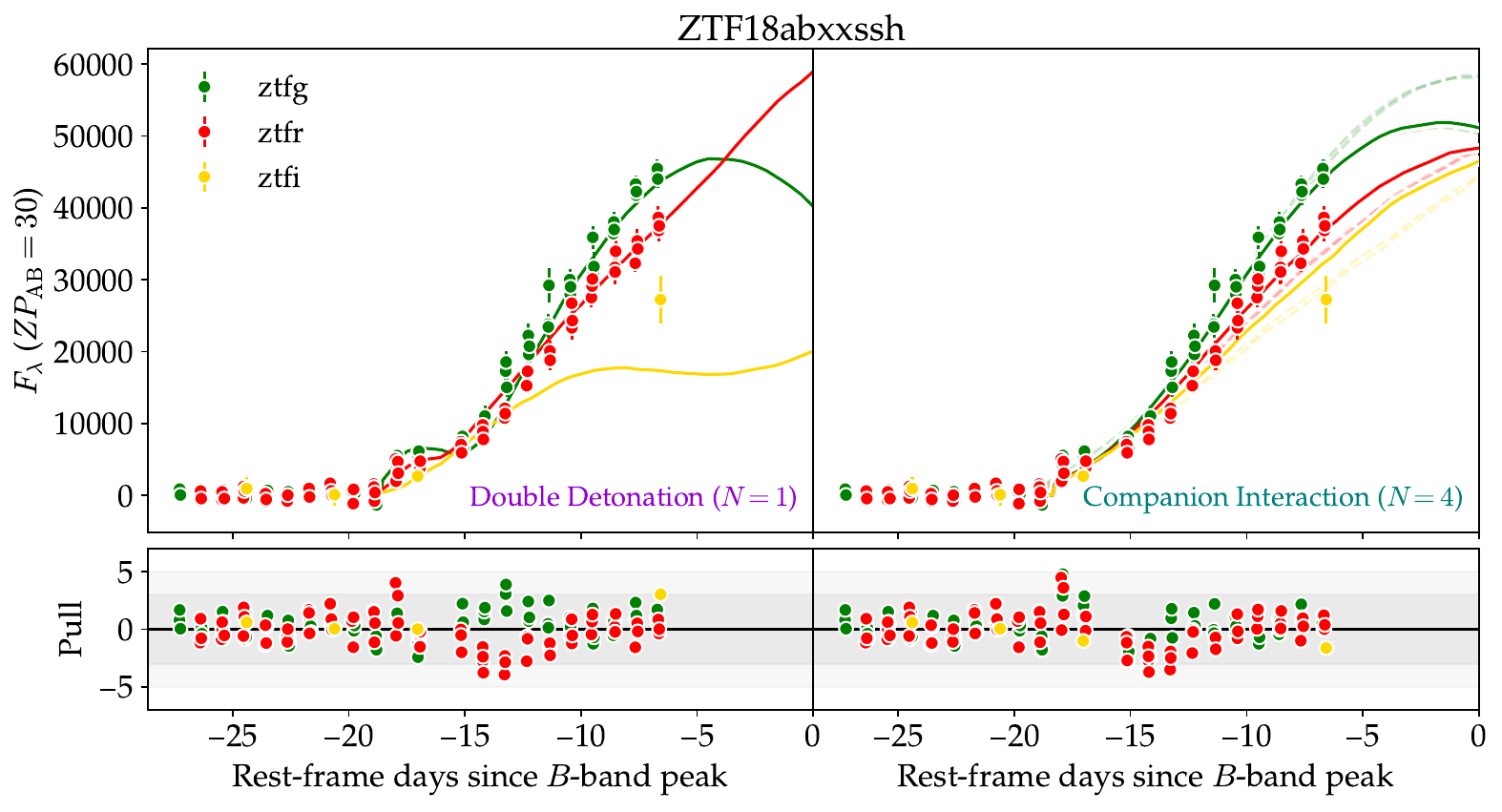}
    \caption{ZTF \textit{gri} rest-frame light curves of \ztfname. The best-fit models (semi-transparent dashed lines), with the number of models included (N), and the best-fit model (i.e., lowest $\chi^2$; solid lines) are shown for the double-detonation (\textit{left}) and companion-interaction (\textit{right}) models (see Sect.~\ref{subsubsec:dd_ci}). The residuals are with respect to the best-fit model to help visualise whether the models are a good approximation of the observations.}
    \label{fig:turtls_example2}
\end{figure}

We follow a different approach to that of \citetalias{Deckers+2022} to select the best-fit models. For each SN, we first retain only those models that reproduce at least 68\% of the data points within $3\sigma$ over the phase range $\phi \in [t_{\rm{fd}}, t_0]$, ensuring that the selected models provide an adequate representation of the observations. All models originate from the same parent model (DD or CI), only differing in the set of four parameters (described above). Therefore, instead of selecting the best-fit models, we select the population of parameters lying within the 99.97\% ($3\sigma$) confidence interval from the best parameters (i.e., lowest $\chi^2$). The extent of this interval depends on the number of free parameters in the model; both the DD and CI models considered here have four parameters. Note that $\mu$ and \texp are considered nuisance parameters and thus not treated as free parameters. Therefore, we select all the models whose parameters are within $\Delta\chi^2=\chi^2-\chi^2_{\rm{min}}{\sim}16.25$\footnote{We follow Section 16.6.5 from Numerical Recipes \citep{numerical_recipes}.}. An example of the best-fit DD and CI parameters/models for a SN of our sample is shown in Fig.~\ref{fig:turtls_example2}.

\section{Results}
\label{sec:results}

In this section, we present the \sneia identified as \eexsn candidates by the different methods used in the previous section.

\subsection{Models without flux excess}
\label{subsec:models_without_excess}

Early flux excess is expected to occur primarily towards bluer bands. Therefore, we focus on $g$-band observations and require at least two epochs with $>5\sigma$ deviation from the models during the excess window to minimise false positives\footnote{We refer to false positives as those SNe incorrectly identified as \eexsne. However, have in mind that there is no prior knowledge of the true \eexsne. Identifying them is the aim of this work.}. As the duration of the early excess in \snia light curves is not precisely known and the epoch of explosion is difficult to estimate, we test different excess windows for the detection of the early flux excess. The optimal excess window is chosen independently for each method (see below), considering that each method has its own caveats (see Sect.~\ref{subsec:methods_comparison}).

With the SALT2 method (Sect.~\ref{subsec:salt2}), we tried different excess windows ranging from 1 to 5 days. The excess window that maximises the number of identified \eexsne (15 objects) is 4\,d, while a 5\,d window does not identify more candidates. These 15 candidates are shown in Fig.~\ref{fig:salt2_eexsne}. From visual inspection of the light curves and SALT2 models, we believe most of them to be viable \eexsn candidates. We also note that around $5\%$ of the SNe have their flux over-predicted by the model around $\phi\sim15$\,d, which was expected as discussed above \citep{Rigault+2025b}.

Using the power-law method (Sect.~\ref{subsec:powerlaw}), we tried excess windows ranging from 1 to 4 days. In this case, a window of 3 days seems to maximise the number of \eexsn candidates, although some likely false positives are identified from visual inspection of the fits (see Fig.~\ref{fig:powerlaw_eexsne}). When the excess window is too large, the few data points remaining are not enough to properly constrain the power-law fit, which results in a somewhat linear rise instead of a smooth, physical-looking curve, causing some of these false positives. We note that the main limitation of this method is that it tends to under-predict the SN flux, depending on the data coverage.

In the case of the modified power-law method (Sect.~\ref{subsec:powerlaw}), we see that most excess windows produce a suspiciously high number of candidates. Although several fits look reasonable, this model seems to vastly under-predict the underlying early SN flux in a systematic manner (see Fig.~\ref{fig:modified_powerlaw_eexsne}). We select 2 days as the optimal excess window to be conservative with the identification of \eexsn candidates (27 objects) as there are only four objects using a 1 day excess window. Note that we do not rely on a single method for the identification of the best candidates (Sect.~\ref{subsec:best_candidates}).

Finally, for the \nickel-distribution models (Sect.~\ref{subsubsec:ni_dist}), we only selected the best-fit model for each SN for identifying early excesses and tested the excess window range from 1 to 4 days. The longer the excess window, the larger the bias in the fits at early times (a caveat with this method), which translates to larger number of false positives, as with the modified power law. Visual inspection of the fits with 3 and 4 days excess windows confirms this. So, we chose the 2\,d window as the optimal one, identifying a conservative number of 34 \eexsne (see Fig.~\ref{fig:nickel_dist_eexsne}), as with 1\,d only nine candidates are identified. These candidates with the best-fit models and parameters are shown in Fig.~\ref{fig:nickel_dist_eexsne}.

In summary, using the optimal excess windows, we identified 15, 8, 27 and 34 possible \eexsn candidates with the SALT2, power-law, modified power-law and \nickel-distribution methods, respectively (see Table~\ref{tab:excess}). Note that not all the \sneia have the data-coverage required by all four methods. Additionally, note that the optimal excess window largely depends on the method and average cadence of the sample.

\begin{table}[h]
\caption{\eexsne identified by different methods for different excess windows.}
\centering
\begin{threeparttable}

\setlength{\tabcolsep}{2pt}

\begin{tabular}{lccc}
\hline
Method & Number of & Excess  & Section \\
 & \eexsne & window (d) &  \\
\hline
SALT2 & 1,2,10,\textbf{15},14 & 1,2,3,\textbf{4},5 & Sect.~\ref{subsec:salt2} \\
Power-law (PL) & 1,4,\textbf{8},7 & 1,2,\textbf{3},4 & Sect.~\ref{subsec:powerlaw} \\
Modified PL & 4,\textbf{27},61,66 & 1,\textbf{2},3,4 & Sect.~\ref{subsec:powerlaw} \\
\nickel-dist. & 9,\textbf{34},93,138 & 1,\textbf{2},3,4 & Sect.~\ref{subsubsec:ni_dist} \\
\hline
\end{tabular}

\begin{tablenotes}
\item \textbf{Notes.} The optimal excess windows chosen, shown in bold face, are explained in Sect.~\ref{subsec:models_without_excess}.
\end{tablenotes}

\end{threeparttable}
\label{tab:excess}

\end{table}

From the \eexsn candidates identified by the different methods, there are 61 unique objects, from which 17 are identified by at least two of the methods (see Table~\ref{tab:eexsne_common}). From these, SNe ZTF18abmwnov (2018fiw; normal subtype) and ZTF18abxxssh (2018gvj; normal subtype) are identified by all four methods, while SN ZTF20abnrdse (2020qkw; a SN Ia with unknown subtype) is identified by all except the power-law method. In the case of SN ZTF18abxxssh, it can be clearly seen that the first two detections are in excess with respect to all four models, with the last non-detection just $\mathord{\sim}1$\,d before first detection (see Fig.~\ref{fig:salt2_example}, \ref{fig:powerlaw_example}, \ref{fig:turtls_example}). This prominent excess is much brighter than those seen in other candidates. SN ZTF18abmwnov also shows a very clear excess that last ${\sim}$4 days, with the last non-detection just $\mathord{\sim}1$\,d before first detection as well (see Fig.~\ref{fig:salt2_eexsne}, \ref{fig:powerlaw_eexsne}, \ref{fig:modified_powerlaw_eexsne} and \ref{fig:nickel_dist_eexsne}). Surprisingly, this SN does not pass the \texttt{fitquality} cut (\texttt{fitprob}${\sim}10^{-10}$, i.e. $< 10^{-7}$; \citealt{Rigault+2025a}). The early excess of SN ZTF20abnrdse has a duration of ${\sim}2$ days, reaching ${\sim}17$\% of $g$-band peak flux, much brighter than the former two SNe which reach ${\sim}10$\%. The latter SN does not pass the \texttt{fitquality} cut either (\texttt{fitprob}${\sim}10^{-10}$).

\begin{table}[h]
\caption{Best \eexsne candidates identified by more than one method from Sect.~\ref{subsec:models_without_excess}.
}
\centering
\begin{threeparttable}

\setlength{\tabcolsep}{0.5pt}

\begin{tabular}{lcccccc}
\hline
ZTF name & IAU & Sect.\ref{subsec:models_without_excess} & Sect.\ref{subsec:models_with_excess} & Subtype & $z$ & \texttt{fqf=1} \\
\hline
18abfhryc & 18dhw & PL,MPL & DD,CI & norm & 0.032 & True \\
18abmwnov & 18fiw & S,PL,MPL,ND & CI & norm & 0.048 & False \\
18abspqsn & 18gdl & S,PL &  & 03fg & 0.036 & True \\
18abtfvsk & 18fop & MPL,ND & DD,CI & norm & 0.021 & True \\
18abxxssh & 18gvj & S,PL,MPL,ND & DD & norm & 0.066 & True \\
18acxyarg & 18kij & MPL,ND &  & norm & 0.041 & True \\
18acybdar & 18kku & MPL,ND & DD,CI & norm & 0.064 & True \\
19aawmqtd & 19gwg & S,PL,ND & DD & 91T & 0.056 & True \\
19aayfaum & 19hez & PL,MPL & CI & norm & 0.024 & True \\
19abqhikf & 19nvv & MPL,ND &  & 91T & 0.032 & True \\
19aclofmz & 19uaf & MPL,ND & CI & norm & 0.031 & True \\
20aahkgcz & 20axk & PL,MPL & CI & norm & 0.041 & True \\
20aajjxwc & 20btv & PL,MPL & DD & 91T & 0.060 & True \\
20aaqpxtm & 20dgc & S,MPL & DD & 91T & 0.052 & True \\
20aazquhc & 20kbh & MPL,ND &  & 91T & 0.034 & True \\
20abeywdn & 20mon & MPL,ND &  & norm & 0.025 & True \\
20abnrdse & 20qkw & S,MPL,ND & DD & unknown & 0.072 & False \\
\hline
\end{tabular}

\begin{tablenotes}
 \item \textbf{Notes.} S: SALT2, PL: Power Law, MPL: Modified Power Law, ND: \nickel-Distribution, DD: Double Detonation, CI: Companion Interaction. SN names have been shorten. The methods from Sect.~\ref{subsec:models_with_excess} are also shown for completeness. Redshift is approximated to three decimal places. \texttt{fqf=1} is the abbreviation of \texttt{fitquality\_flag=1}.
\end{tablenotes}

\end{threeparttable}
\label{tab:eexsne_common}

\end{table}

\subsection{Models with flux excess}
\label{subsec:models_with_excess}

As a reminder for the reader, DD and CI models have both light curves with and without early excess. In this work, we treat the DD and CI models independently as they were built based on different physical assumptions and parameters.

To identify \eexsne, we calculate the fraction between the best-fit models (best-fit parameters; see Sect.~\ref{subsubsec:dd_ci}) with an early excess ($N_{\rm EE}^{\rm Best-fit}$) and the total amount of available models with an early excess ($N_{\rm EE}^{\rm Total}$), $n_{\rm EE}$, and do the same for the models without an early excess, $n_{\rm No-EE}$. Then, we calculate the probability for a \sneia of having an early excess by calculating the fraction $n_{\rm EE}/(n_{\rm EE} + n_{\rm No-EE})$. A value of $0$ means that no model with early excess was within $\Delta\chi^2 \leq16.25$ of the best-fit model (\chinorm$=1$), while a value of $1$ means that solely models with early excess are selected. As a reminder, only models where 68\% of the data in the range $\in [t_{\rm{fd}}, t_{\rm{fd}}+5]$\,d falls within $3\sigma$ of them are considered. This removes any models that are not a good match to the observations. 

\begin{figure}[h!]
    \includegraphics[width=\columnwidth]{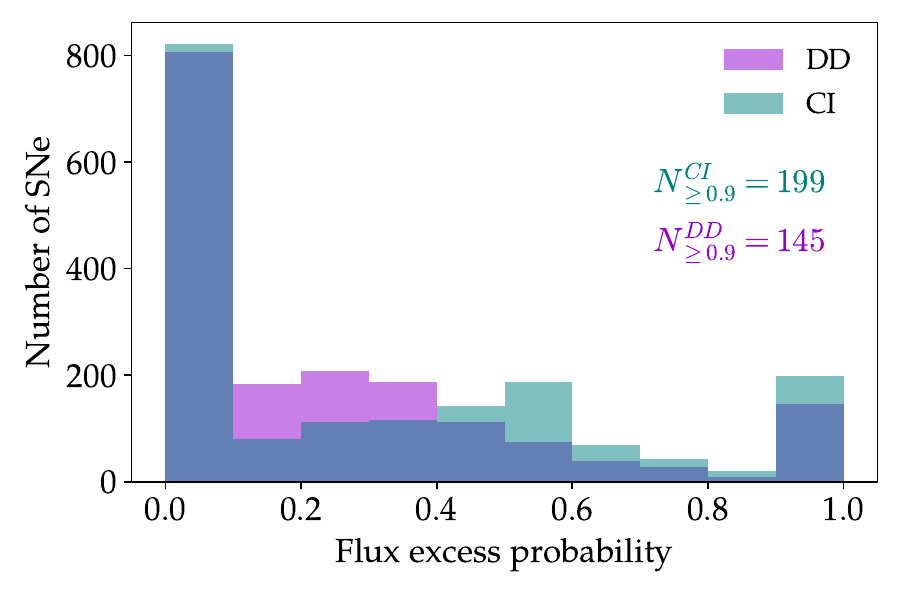}
    \caption{Flux excess probability for the \sneia in our sample, using the \textcolor{Plum}{Double Detonation} (DD) and \textcolor{teal}{Companion Interaction} (CI) models (see Sect.~\ref{subsubsec:dd_ci}). \eexsne are those with a probability $\geq0.9$. Their numbers are shown in the plot ($N_{\geq0.9}$). Sect.~\ref{subsec:models_with_excess} explains how the probability is calculated.}
    \label{fig:excess_probability}
\end{figure}

The probability distributions for our sample of \sneia are shown in Fig.~\ref{fig:excess_probability}. These distributions show a clear difference in the number of \sneia split at a probability of $0.9$. Therefore, we decided to consider \eexsn candidates as those with a probability higher that $0.9$. Although the choice is somewhat arbitrary, it allows us to do a relative comparison between the DD and CI models. 

Using the DD and CI models, we find 145 and 199 \sneia, respectively, with a probability of being \eexsne above $0.9$. This includes models that show `bumps' and `shoulders' in the early light curves that are both classified as early excesses. Moreover, data not only around early epochs are considered for the fits, but also data up to peak flux. Therefore, the preference for these models with early excesses may be driven by a better overall match in light curve shape than specifically the presence of an early excess. This results in a relative large number of \eexsn candidates found by comparison with these models. There are 27 \eexsn candidates identified with both DD and CI models.

\section{Discussion}
\label{sec:discussion}

In this section, we discuss the results found with the different methods used in this work and the possible implications.

\subsection{Methods comparison}
\label{subsec:methods_comparison}

Considering only deviations from canonical rise models (i.e. without early flux excess), we identify 15, 8, 27 and 34 \eexsn candidates using the SALT2, power-law, modified power-law and \nickel-distribution methods, respectively. Note, however, that there are several false positives among these, particularly for the latter two.

The advantage of the SALT2 method is its simplicity and the use of a light-curve fitter tailored to \sneia. However, SALT2 suffers from a limited training set at very early epochs ($\phi \lesssim-15$\,d), where some extrapolation of the model is required, possibly introducing biases \citep[e.g.][]{Rigault+2025b}. Although we have taken some measures to limit these biases, such as checking for agreement between the model and observations outside the excess window, they might not completely avoid them. Additionally, this method requires data after peak to produce an accurate light-curve fit.

\begin{figure}[h!]
    \includegraphics[width=\columnwidth]{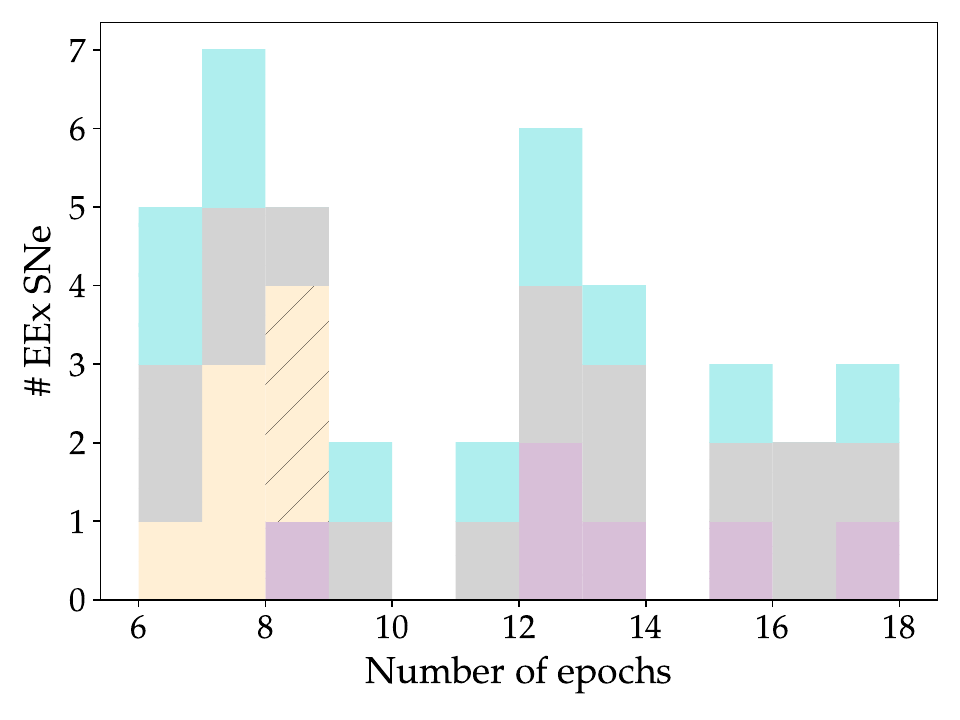}
    \includegraphics[width=\columnwidth]{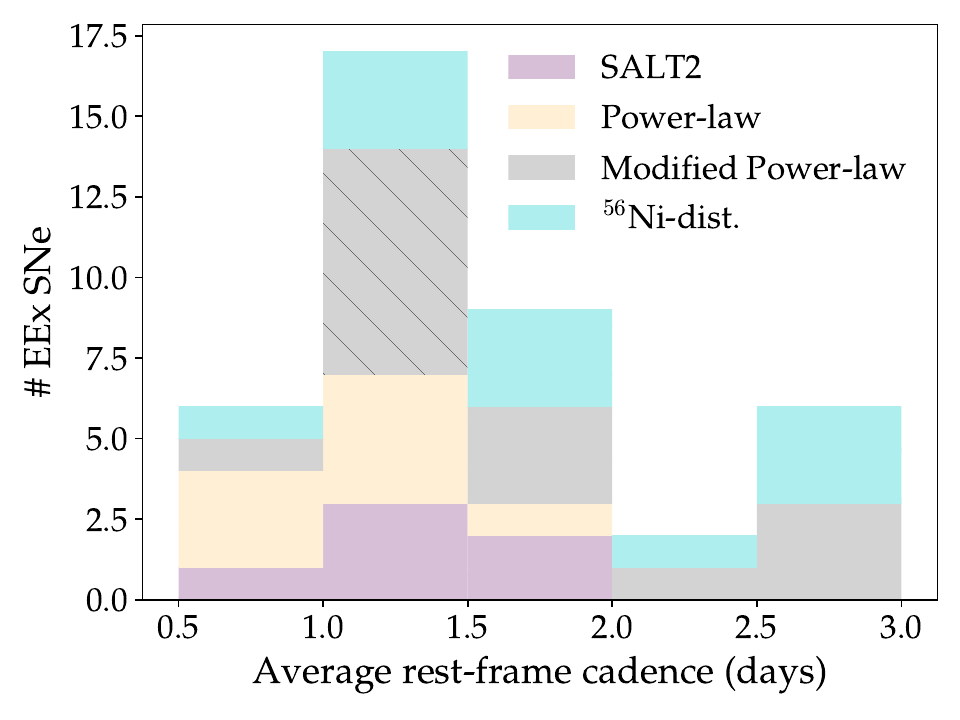}
    \caption{Number of \textit{g}-band epochs (unique nights; \textit{top}) and average rest-frame cadence (\textit{bottom}) during the light-curve rise of \eexsn candidates in this work. The light-curve rise is considered between \tfd and \tmax. The median rest-frame cadence of the \eexsne is $\mathord{\sim}1.5$\,d. Note that the power-law method actually uses data up to the time of 50\% of \g-band peak flux (Sect.~\ref{subsec:powerlaw}), which translate to a lower number of epochs in comparison to the other methods.}
    \label{fig:rise_and_cadence}
\end{figure}

The advantage of the power-law method also lies in its simplicity and ability to describe the early light-curve rise of \sneia in a relatively accurate manner \citep[e.g.][]{Nugent+2011, Miller+2020}. However, this method is greatly limited by the coverage, requiring high cadence, as can be seen in the top panel of Fig.~\ref{fig:rise_and_cadence}. The constraint on the time range used for this method, i.e. data up to the time of 50\% of \g-band peak flux (Sect.~\ref{subsec:powerlaw}), is the main reason why fewer \eexsne are identified, compared to the other methods, although most seem good candidates. Long data gaps can cause unphysical description of the light-curve rise, closer to a straight line instead of a smooth curve.

The modified power-law method has similar advantages as the simple power law, although without the same limitations in data coverage. It does also introduce additional free parameters in the fitting. Nonetheless, given the large number of \eexsn candidates found of which most are likely false positives, it seems to systematically under-predict the SN flux at early epochs as well. Importantly, not all candidates identified by the power-law method are flagged by the modified power law.

The \nickel-distribution method benefits from having an astrophysics-based model. However, it is limited by the requirement of very early data to avoid biases \citep[][]{Magee+2020}. The inclusion of baseline data before first detection does not seem to alleviate the issue of under-predicting early-time flux. Additionally, the modelled light curves might not necessarily be an accurate representation of the observed light curves. The available parameter space used for creating these models is another limitation of this method. A finer grid of these parameters would be desired, but the computing time increases proportionally.

Despite the differences between the methods, they all need high cadence observations (mostly $\lesssim2$\,d) and at least around six epochs during the rise (see Fig.~\ref{fig:rise_and_cadence}). We note that the power-law method in particular seems to require higher cadence than the other two given its limitation in the phase range used, as discussed above. Given the substantial variability in number of epochs during the rise (top panel of Fig.~\ref{fig:rise_and_cadence}), we conclude that aside from the high average cadence, lack of long gaps (e.g., $\gtrsim3$\,d) in the observations is ideal for unequivocally identifying early excesses. Unfortunately, these observing gaps are typically unavoidable with wide-field surveys like ZTF.

The SALT2, modified power-law and \nickel-distribution methods need data at least up to around peak brightness (even beyond for SALT2) to obtain satisfactory results. This makes it really difficult to identify \eexsne at very early epochs when the SN is still rising. On the contrary, the power-law method is probably the best method for identifying \eexsne in almost \lq real time', although it still needs data during the rise (several days after the explosion) for this.

The main advantage of the DD and CI methods, as with the \nickel-distribution method, is that they benefit from having an astrophysics-based model. Furthermore, they both include models with and without early excess, which in principle makes them more representative of real \sneia.

The CI model light curves peak, on average, around 18.9 and 19.8 days after explosion in \g and \r bands, respectively. However, the \g-band light curve of the models with $s=9.7$ peak later (20.4 days after explosion) compare to those with $s=3$ (17.5 days after explosion), although with a large spread in values and little difference in the \r band. This implies that the models with early excess tend to have longer rise times. Surprisingly, we do not see that the \eexsne identified by this model have higher stretch than the rest of the SNe (see Sect.~\ref{subsec:best_candidates} and Fig.~\ref{fig:distributions_CI}). Other parameters do not seem to largely alter the rise time.

The DD model \g-band light curves peak, on average, around 16.1 days after explosion, earlier than those of the CI model. The \r-band light curves peak around 19.3 days after explosion, but rise more linearly compared to the CI model (see Fig.~\ref{fig:turtls_example2}). Different parameters alter the rise time in the \g band, producing a large range of values, whilst the \r-band rise time remains mostly unchanged. Interestingly, we do not see any clear trend between rise time and the burning shell fraction or with the dominant burning product. The shorter rise times in the \g band and the more linear rise in the \r band represent a somewhat less realistic behaviour of the light curves, which could potentially explain the lower number of \eexsn candidates found with the DD method. 
We note that the physical parameters that affect the early excess of the models also affect the overall shape of the light curves. Hence, we cannot say a priori whether the early excess of the candidates is well described by the models.

We note that several of the best \eexsn candidates (Sect.~\ref{subsec:best_candidates}) show an excess, at a $5\sigma$ level, in their \r-band light curves as well, depending on the method used. Inspecting their early $(g-r)$ colour evolution (see Fig.~\ref{fig:early_colours}), several of the objects seem to show possible hints of a `red bump', as predicted by double-detonation models \citep[e.g.][]{Bulla+2020, Ni2022, Ni2023}, although this is not completely clear. Further investigation is out of the scope of this work.

\subsection{Comparison with previous works}
\label{subsec:previous_works}

As a test for the methods used in this work, we select our \eexsn candidates from 2018 and compare them against the samples from \cite{Yao+2019}, \cite{Bulla+2020}, \citetalias{Miller+2020}\footnote{\citetalias{Miller+2020} did not identify any \eexsne but we add their work for completeness.}, \citetalias{Deckers+2022} and \cite{Burke+2022}. These are summarised in Table~\ref{tab:2018_sne}. As can be clearly seen, the identification of early excess is highly method dependent. We identified eight \eexsn candidates also identified by other works, of which two are flagged by more than one method: ZTF18abxxssh and ZTF18abfhryc (2018dhw), both of normal subtype. Particularly, there is a general agreement that ZTF18abxxssh shows an early excess in its light curve, as we found above. From the methods in this work, only the CI method did not identify SN ZTF18abxxssh as a \eexsne candidate, possibly because these models do not show such prominent bumps (see Fig.~\ref{fig:turtls_example2}).

From the comparison with the literature sample, we note that despite using similar methods as in \citetalias{Miller+2020} and \citetalias{Deckers+2022}, we found different \eexsn candidates. \citetalias{Miller+2020} did not make a systematic search of \eexsne in their analysis, but we consider their work for completeness. We believe that if the excess window is not excluded from the power-law fit, it is very likely that the model will actually fit through the excess, which could explain the absence of \eexsn candidates in their work (see also \citealt{Liu_Miller2006, Liu+2026}). Comparing with the method from \citetalias{Deckers+2022}, we also believe that the exclusion of the excess window is needed. \cite{Burke+2022} used companion-shocking models assuming a single-degenerate progenitor, while \cite{Bulla+2020} used early $g-r$ colour information to identify \eexsn candidates. 

A caveat to this comparison is that these previous works used `forced' point-spread function photometry, presented in \cite{Yao+2019}, which can differ from the published one (\smith). Although large differences are not expected, they can be significant for these analysis as the methods used are very sensitive to the data quality. Further analysis in this regard falls outside the scope of this work.

\subsection{Best \eexsn candidates}
\label{subsec:best_candidates}

Given the large heterogeneity in the results provided by the different methods, and even within the same method but using slightly different approaches (e.g., \citetalias{Deckers+2022} vs this work), we believe the best \eexsn candidates to be those identified by multiple methods. However, we only consider the methods from Sect.~\ref{subsec:models_without_excess} as those from Sect.~\ref{subsec:models_with_excess} rely on stronger assumptions about the underlying physics. The latter methods are used to set a conservative upper limit on the \eexsne.

Following the discussion above, SN ZTF18abxxssh can be considered a canonical \eexsne as it is identified by most methods. However, one must consider that this is in part due to the prominence of its early excess; a large diversity from `bumps' to `shoulders' is expected, which makes it hard to identify the latter cases. 

To rule out false flux excesses, we inspect the light-curve logs of the best candidates to check whether the observations during the supposed excesses coincide with a different camera chip with respect to the rest of the observations during the light-curve rise. No relation between the excesses and specific chips were found.

From Sect.~\ref{subsec:models_without_excess}, we found 17 objects to be the best \eexsn candidates. Of these, six are of an over-luminous subtype (five 91T-like and one 03fg-like; see Table~\ref{tab:eexsne_common}), one of unknown subtype (SN ZTF20abnrdse), while the rest are all normal. Interestingly, three of the 91T-like SNe are flagged by the DD method and not the CI method, which could potentially imply a preferred explosion mechanism for these over-luminous \sneia. A more in-depth analysis is left for future work. The larger relative fraction of 91T-like and 03fg-like \sneia, compared to normal SNe Ia, is in agreement with the findings of \cite{Jiang+2018}. These could be either due to an intrinsic difference in the explosion mechanisms, where over-luminous \sneia tend to have more prominent bumps, or due to an observational bias, where higher-stretch SNe have more observations during their rise. 
Inspecting the pseudo rise time (Fig.~\ref{fig:bumpy_rise_from_tfd}) and number of epochs during the rise of these \eexsn candidates (Fig.~\ref{fig:rest_rise_and_cadence}), we find no clear correlation. Thus, we believe the former to be the most likely possibility. Future works will help shed light on this matter.

To estimate the rate of \sneia with early excess, we restrain the samples to be volume-limited (i.e., $z \leq0.06$; \citealt{Amenouche+2025}), leaving only 645 SNe. We consider the best candidates to be a lower limit on the number of \eexsne, noting that this selection is restricted to those with prominent excess. Of the 17 SNe, 13 are in the volume-limited sample. In this manner, we retrieve the following minimum rates of \eexsne, divided per subtype: $2\%$ for normal, $11\%$ for 91T-like and $17\%$ for 03fg-like. On the other hand, we consider the candidates identified by the methods from Sect.~\ref{subsec:models_with_excess}, jointly with the best candidates\footnote{Note that these candidates are not necessarily flagged by either of the DD or CI methods (see Table~\ref{tab:eexsne_common}).}, to be an upper limit. Assuming the DD model as the explosion mechanism, we obtain the following maximum rates: $11\%$ for normal, $32\%$ for 91T-like and $17\%$ for 03fg-like (i.e., no 03fg-like candidates were flagged by this method). Assuming the CI model as the explosion mechanism, we obtain: $27\%$ for normal, $13\%$ for 91T-like and $50\%$ for 03fg-like. These values are in agreement with previous works \citep[e.g.][]{Magee+2020, Deckers+2022}, but do depend on the exact method used. It is worth noting that the DD and CI models may be a good representation of normal \sneia but not necessarily those of peculiar subtypes. Also, the population of non-\eexsne differs between both methods.

\subsection{\eexsn candidates: parameter distributions}
\label{subsec:parameters_distributions}

To further analyse the sample of \eexsne, we compare their light-curve parameters and host properties against the rest of \sneia, using the volume-limited sample. We only consider the SNe whose subtypes are in common with those of the best candidates, i.e. normal, 91T-like and 03fg-like SNe. In addition, we remove objects that do not pass the \texttt{fitquality} cut, leaving 559 SNe in total. Note that the same cuts are applied to the non-\eexsn and \eexsn candidates.

\begin{figure*}[ht!]
    \includegraphics[width=0.24\textwidth]{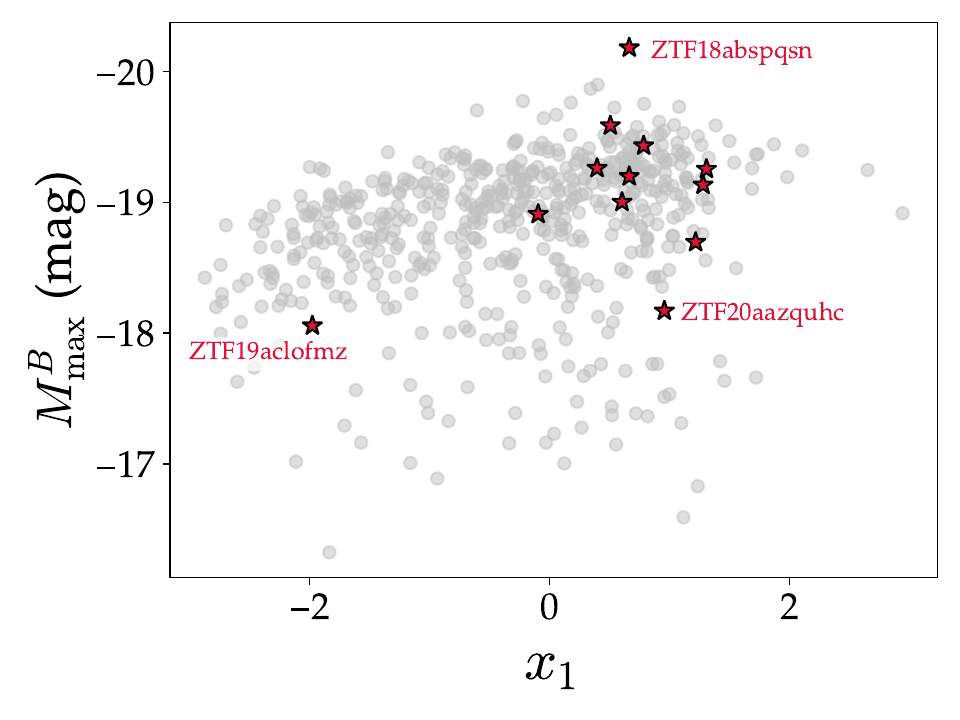}
    \includegraphics[width=0.24\textwidth]{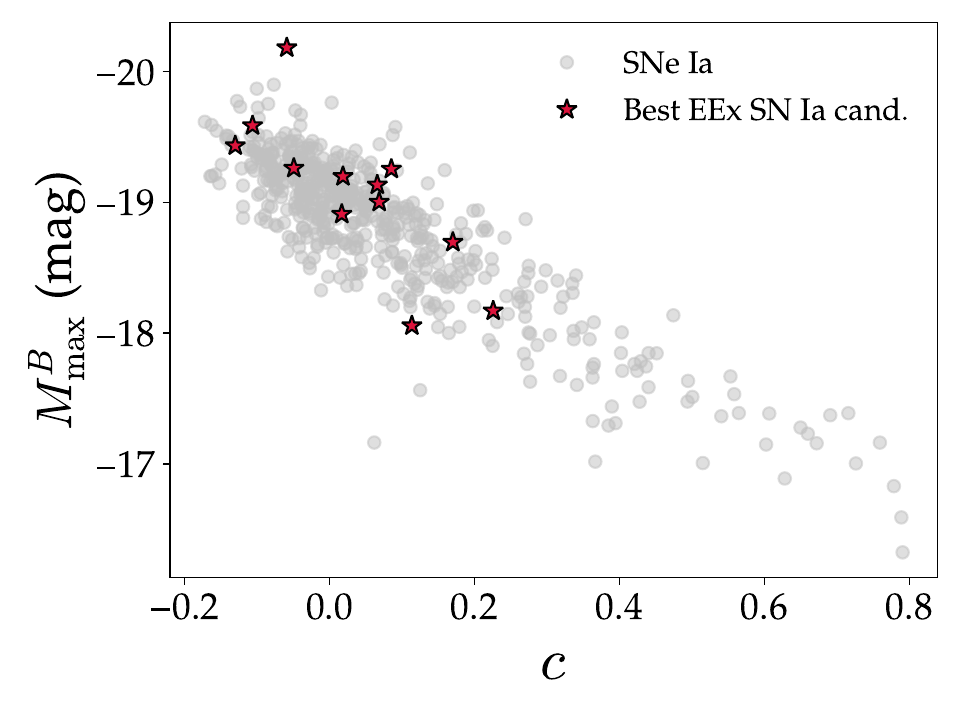}
    \includegraphics[width=0.24\textwidth]{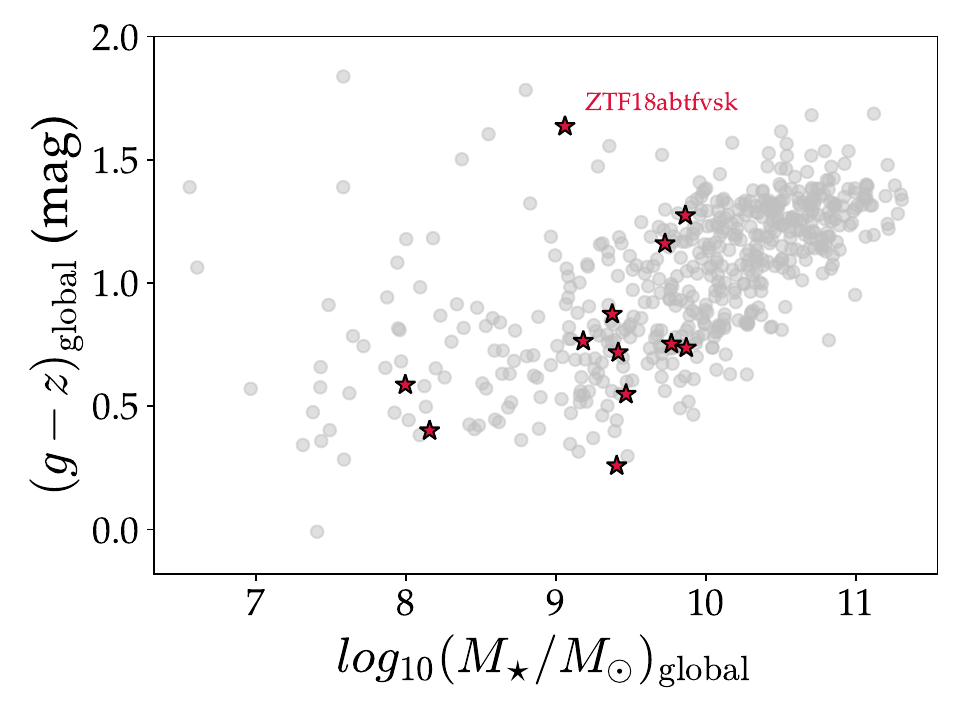}
    \includegraphics[width=0.24\textwidth]{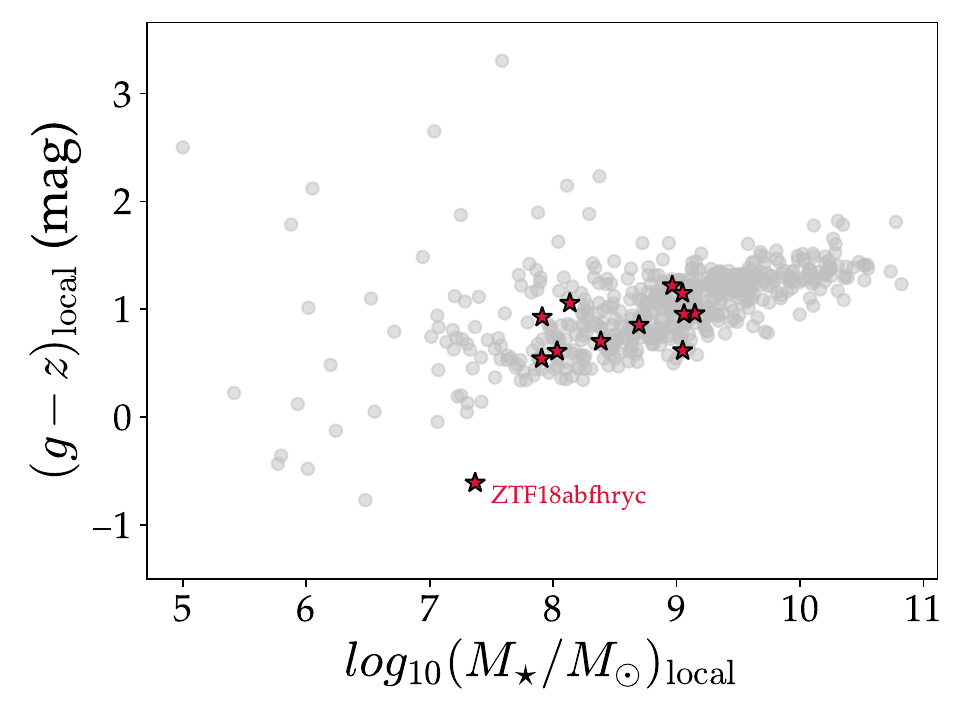}
    \caption{Distributions of light-curve parameters and host properties for the non-\eexsne (assuming the DD model; grey circles) and best \eexsn candidates (crimson stars). Some \eexsn candidates with values outside the bulk of the population are marked. The distributions for all the candidates using the DD and CI methods are shown in Fig.~\ref{fig:distributions_DD} and \ref{fig:distributions_CI}, respectively.}
    \label{fig:parameter_distributions}
\end{figure*}

In Fig.~\ref{fig:parameter_distributions}, we show the distribution of parameters for the best \eexsn candidates (see Fig.~\ref{fig:distributions_DD} and \ref{fig:distributions_CI} for the candidates using the DD and CI methods, respectively). The parameters are the \textit{B}-band peak absolute magnitude ($M^B_{\rm max}$\footnote{A flat $\Lambda$CDM cosmology with $H_0=70$\,\hunits and $\Omega_m=0.3$ is assumed to calculate the distance.}), stretch ($x_1$) and colour at peak ($c$) \citep{Rigault+2025a}. We also show both global and local host properties: the stellar mass, $log_{10}(M_{\star}/M_{\odot})$, and $(g-z)$ rest-frame colour (\smith). The directional light-radius distance distribution, $d_{\rm DLR}$ \citep{Sullivan+2006, Gupta+2016}, is considered as well. All the parameters were collected using the \texttt{ztfidr} package.

By using the two-samples Kolmogorov-Smirnov (KS) test \citep{Kolmogorov1993, Smirnov1939}, we check whether the parameter distributions of \eexsn candidates are compared to the rest of \sneia. We consider the distributions to be significantly different if the p-value (p$_{\rm KS}$) is lower than 0.05. For the best \eexsn candidates, we find that they have higher $x_1$ on average than the comparison sample. This trend is in line with having a larger number of over-luminous subtypes. \cite{Wu+2025} found that \eexsne have, on average, longer rise times and brighter peak magnitudes, but we do not find evidence for the latter trend. More recently, \cite{Wu+2026} found a very similar trend with $x_1$ for three out of 16 \sneia identified as \eexsne using the 2.5m Wide Field Survey Telescope \citep{Hu+2022}. Our best candidates also prefer lower-mass (both global and local) and bluer hosts (global only) (see Fig.~\ref{fig:parameter_distributions} and \ref{fig:distributions_DD}). These trends cannot be explained by the over-luminous subtype \citep{Dimitriadis+2025} but could be explained by high-stretch SNe preferring these type of environments \citep{Ginolin_2025_x1}, connecting the excess to younger stellar populations.

Although no significant difference is observed in the colour of the best candidates, we note that none has $c>0.3$. The apparent absence of extinction observed in these objects could possibly be connected to young stellar populations as well. However, dust could also attenuate the early excess making it undetectable in highly extincted SNe.

When we consider the \eexsn candidates using the DD method (Fig.~\ref{fig:distributions_DD}), they are brighter and bluer on average than non-\eexsne. This is probably due to the relatively large fraction of over-luminous \eexsn candidates compared to the normal ones (for every four normal candidates, there is one over-luminous candidate). This is somewhat unexpected given some of the properties of these models, such as their intrinsically red colours \citep[][]{Magee+2021}, although these models can reach high peak brightness as well (see Sect.~\ref{subsec:ee_models}). These candidates also seem to prefer lower-mass hosts (local values). The local host colour and $d_{\rm{DLR}}$ distributions differ from the non-\eexsne as well, according to the p-value, where extreme values are not favoured by the \eexsn candidates.

In the case of the CI method (Fig.~\ref{fig:distributions_CI}), we find that the \eexsn candidates are fainter and redder than the comparison sample on average. With this method, we find that, of all the candidates, approximately 80\% are normal, 12\% are under-luminous and 3\% are over-luminous. These models could be better at describing the light curves of some under-luminous SNe. Interestingly, we do not find any significant difference in $x_1$ or other host properties compared to the non-\eexsne.

Notably, the objects and trends observed in the light-curve and host-galaxy properties differ between the best candidates and those identified by the DD and CI methods. This may indicate that the models are not an accurate representation of the whole \eexsn population, although one must also consider the limited parameter space used in study.

\subsection{Early-Excess Models: parameter distributions}
\label{subsec:ee_models}

The DD and CI models are both viable scenarios for describing \sneia explosions. In this section, we analyse the best-fit models of the \eexsn candidates to help constrain the explosion mechanism of \sneia.

One of the first things that stand out is the relatively few objects in common between the \eexsne identified with the DD and CI methods: only 21 SNe. We believe the main reason behind this to be the difference in light-curve shapes of the models. As both models have different physical assumption, different outputs are expected. For instance, as discussed in Sect.~\ref{subsec:models_with_excess}, the DD models have \g-band light curves that peak earlier than expected and with \r-band light curves with a somewhat linear rise. On the other hand, CI models show more physical-looking light curves in general. From the peculiar \eexsn candidates, it is clear that the DD model prefers over-luminous SNe, while the CI model prefers under-luminous ones.

\begin{figure}[h!]
    \includegraphics[width=\columnwidth]{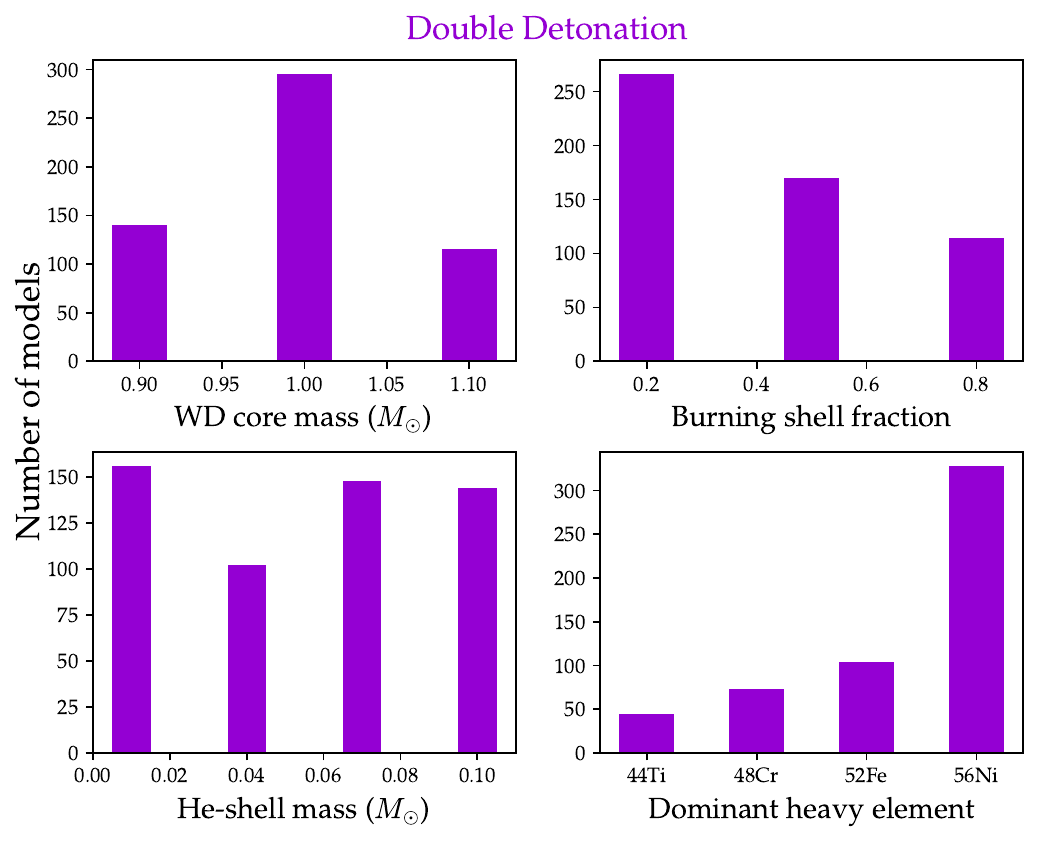}
    \caption{Parameter distributions for the best-fit double-detonation models for \eexsne identified in Sect.~\ref{subsec:models_with_excess}.}
    \label{fig:parameter_distributions_DoubleDetonation}
\end{figure}

\begin{figure}[h!]
    \includegraphics[width=\columnwidth]{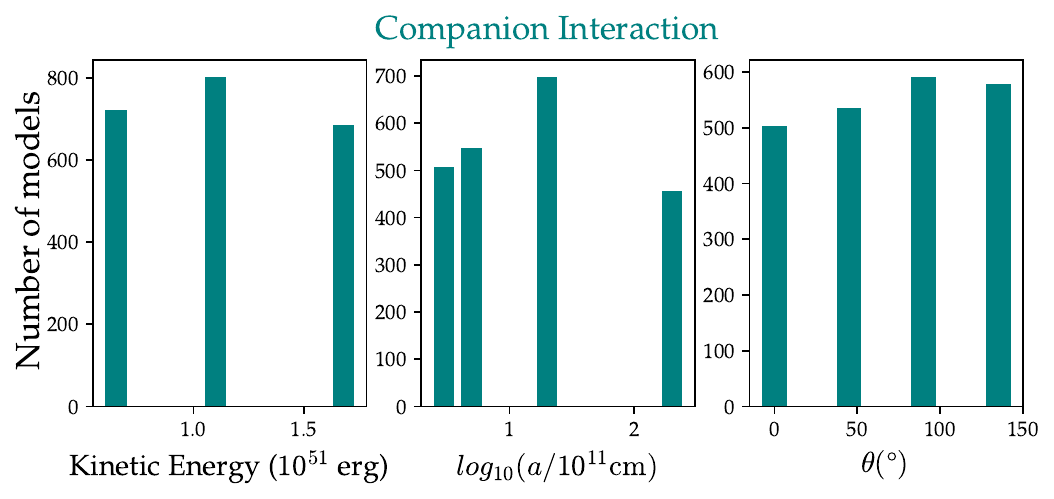}
    \caption{Parameter distributions for the best-fit companion-interaction models for \eexsne identified in Sect.~\ref{subsec:models_with_excess}.}
    \label{fig:parameter_distributions_CompanionInteraction}
\end{figure}

Treating the DD and CI models independently, and assuming the number of \eexsne identified by each of these to be true, we can study their physical properties. The parameter distribution for the best-fit models from Sect.~\ref{subsec:models_with_excess}, for both the DD and CI models, are shown in Figs.~\ref{fig:parameter_distributions_DoubleDetonation} and \ref{fig:parameter_distributions_CompanionInteraction}, respectively. The distributions for the non-\eexsne can be found in Fig.~\ref{fig:parameter_distributions_nonEExSNe_DoubleDetonation} and \ref{fig:parameter_distributions_nonEExSNe_CompanionInteraction}. Note that some of the models can contribute more than once if they are a good match for multiple \sneia.

It can be seen that, considering the available parameter space, \eexsne have a preference for DD models with WD core mass $1$\,\msun, a lower fraction of shell burning ($20\%$) and heavier He-burning products (\nickel). No He-shell mass in particular is significantly preferred. Non-\eexsne on the other hand, focusing only on models without excess, show a preference for $1.0-1.1$\,\msun WD core mass, and the obvious preference for lighter He-burning products (e.g., $^{32}$S). They also seem to have a slight preference towards lower He-shell masses. 

The DD models span a wide range in peak brightness, roughly from ${\sim-}18$ to $-20$\,mag in the $g$ band, mainly driven by the WD core mass (brighter models have higher values). However, this also depends on the exact combination of parameters. In general, most parameters affect the entire evolution of the light curves, except for the He-burning products that mainly affect the first few days. By studying the flux excess of the models, we see that a lower WD core mass tends to produce a more prominent and sharply defined early excess. Reducing the He-shell mass generally shortens the duration of this excess but produces a noticeable bump, whereas higher shell masses can prolong it and even merge it with the main rise, forming a noticeable \lq shoulder' in the light curve. A lower burning shell fraction results in a fainter and slightly shorter bump compared to higher fractions. Additionally, heavier dominant products from He burning typically generate a stronger and more distinct early excess. We note that some of these properties have somewhat opposite effects but, in general, \eexsn candidates have a preference for relatively long-duration, sharply defined early excess.

In the case of the CI model, \eexsn candidates do not seem to show a preference towards any set of parameters. If the true population of \sneia were to present a continuum in early excess properties, one could potentially expect this uniform distribution in physical parameters. However, for the viewing angle in particular, it would be less likely to see the explosion along the line of sight of the shocked region ($\theta=0^{\circ}$) than at higher inclinations (e.g., $\theta=90^{\circ}$), which is not clearly seen in Fig.~\ref{fig:parameter_distributions_CompanionInteraction}.
Non-\eexsne, on the other hand, only show a preference towards larger kinetic energy ($1.65\times10^{51}$\,erg; Fig.~\ref{fig:parameter_distributions_nonEExSNe_CompanionInteraction}).

The CI models span a narrower range in peak brightness, compared to the DD models, going roughly from ${\sim}-18.8$ to $-19.5$\,mag in the $g$-band. Note that this distribution does not closely agree with the observed \textit{B}-band peak absolute magnitude distribution (Fig.~\ref{fig:distributions_CI}; particularly the faint end) as the models do not necessarily agree with observations around light-curve peak (Sect.~\ref{subsec:models_with_excess}). The CI models also have bluer and a narrower range in colour around peak. Parameters like the viewing angle ($\theta$) and the companion separation ($a$) mainly affect the early light-curve evolution, while the kinetic energy affects later epochs up to peak. Therefore, distinguishing between an explosion with no companion interaction and one with companion interaction signal at high inclination (e.g., $\theta{\sim}180^{\circ}$) might not be possible. As previously discussed, models with more extended nickel distribution, i.e. lower $s$ values, rise more steeply with shorter rise times to peak compared to those with more compact nickel distributions. Studying the early excess we see that smaller companion separations tend to create more distinct but shallower early excesses, while larger separations produce a brighter excess that blends into the light curve as a shoulder. Lower kinetic energies generally yield more clear bumps while larger values produce more shoulder-like excess. Viewing angles oriented away from the shocked region also diminish the prominence of the excess, replacing a sharp bump with a shoulder. As with the DD model, some of these properties have opposite effects. The lack of general trends lead us to think that \eexsn candidates may show a continuum in the shape of the early excess.

In general, the CI models show a finer grid in terms of the early-excess shape, while the DD models show a wider distribution in the strength and duration of the excess. However, as was already mentioned, the fits are made with data up to peak, where the effect on the light-curve shape produced by different parameters is harder to distinguish. If the early excess were to more commonly manifest as a shoulder instead of a clear bump, it could possibly explain the larger number of \eexsne detected with the CI model. SN 2018oh \citep{Dimitriadis+2019, Shappee+2019} is a clear example of this (see \citealt{Fausnaugh+2021} for other \sneia with early coverage as well). However, as data up to peak flux is used for fitting the DD and CI models, is difficult to say whether they are a good match to the early excess. The \eexsn candidates show strong trends with some parameters of the best-fit DD models (e.g., burning shell fraction, dominant heavy element), while no such trends are clearly seen with the CI model parameters. Unfortunately, it is difficult to draw any strong conclusions on whether one of the models is preferred over the other to explain \sneia explosions.

\section{Conclusions}
\label{sec:conclusions}

In this work, we tested different methods for the identification of early flux excess in the light curves of \sneia (i.e. \eexsne) from the ZTF DR2 sample. We used two main approaches: i) finding deviations between early-time observations and models without excess, and ii) agreement between observations and models with excess. For the former, we use SALT2, power-law, modified power-law, and \nickel-distribution (TURTLS) models, while for the latter we use the double-detonation (DD) and companion-interaction (CI) TURTLS models.

With the first approach, we define the \lq excess window', where early flux excess is expected and data in this range is not included in the fits. We tested different lengths and selected the optimal window for identifying \eexsn candidates. We found different number of objects with the different methods. For the second approach, we identify \eexsne by selecting objects with a large relative fraction ($>0.9$) of best-fit models with early excess over models without early excess. We found 145 and 199 \eexsn candidates with the DD and CI models, respectively. 

Few \eexsne were found in common between the different methods, evidencing the dependence on the specific method used. This is also apparent when we compare our analysis with previous works on \sneia from the ZTF 2018 sample. We identify 17 best \eexsn candidates as those flagged by more than one method. We believe that no single method is, in general, enough to accurately identify early excess, unless an object presents a very prominent bump. We find a general agreement that ZTF18abxxssh, ZTF18abmwnov, ZTF20abnrdse are \eexsne, mainly due to the prominent excess they present, which can be used as archetypes in future studies with upcoming data releases of ZTF \sneia. Their early excess range from ${\sim}10$ to $17$\% of peak flux (${\sim-}16.2$ to ${-}17.0$\,mag) and last ${\lesssim}4$ rest-frame days.
Cases like ZTF19adcecwu \citep[2019yvq, of 02es-like subtype;][]{Miller+2020b, Burke+2021}, which show a clear single epoch with early flux excess, are not identified by our analysis given that we require at least two epochs with excess to avoid false positives. 

By constraining our sub-sample to be volume-limited ($z<0.06$) and considering the best \eexsn candidates as a lower limit and all possible candidates as an upper limit, we estimate that roughly up to $\mathord{\sim}25\%$ of \sneia are expected to be \eexsne, assuming the CI model, in agreement with previous works. Assuming the DD model, one of the leading models for explaining \sneia, a smaller rate of $\lesssim13\%$ is found. Both upper limits rely on assumptions on the physical mechanism explaining \sneia and identify different populations of \eexsne/non-\eexsne (Sect.~\ref{subsec:models_with_excess}), so the actual rate might differ.

When analysing the sample of best \eexsn candidates, we found a larger relative fraction of over-luminous subtypes \sneia (five 91T-like and one 03fg-like), compared to normal ones (12; not including one of unknown subtype). When studying the light-curve parameters of the sample, we see a significant difference in $x_1$ parameter, where the best \eexsn candidates have larger stretch on average. Although this could be possibly caused by an observational bias, as these objects would have more data in their rise and would therefore be easier to inspect for early excess, we find no clear evidence. These SNe also show lower-mass and bluer hosts, which potentially links them to younger stellar populations as has been suggested from previous smaller samples \citep[e.g.][]{Deckers+2022}.

We also study the distribution of parameters, within the available parameter space, of the best-fit DD and CI models from \eexsne. For the DD models, we find a preference for a WD core mass around $1$\,\msun, heavier He-burning products, and lower fraction of shell burning. For the CI models, we do not find any trends. We cannot draw any strong conclusions on whether one of the models is preferred over the other as the explosion mechanism of \sneia.

In general, each of the methods used has different advantages and disadvantages. For instance, the SALT2 method is simple to implement but needs data after peak. The power-law method suffers from low data numbers given the narrow phase range needed, but is probably the only method (tested in this work) that can be used in almost \lq real time' for the identification of \eexsne. The modified power law does not suffer from the data coverage limitation as the simple power law daos, but it tends to under-predict the SN flux at early epochs. The TURTLS models have the benefit of being astrophysics-based, however they are limited by the parameter spaced used for building them. Emulators may offer a significantly broader parameter space for theoretical models while requiring substantially less computational time \citep[e.g.][]{Magee+2024}, which can be part of future work.

The Vera C. Rubin observatory Legacy Survey of Space and Time \citep[LSST;][]{Ivezic+2019} will start its main survey in 2026 and will discover up to millions of SNe in its 10-year survey. The latest proposed observing strategy for the LSST deep drilling fields\footnote{\lq Ocean' DDF survey strategy: \url{https://community.lsst.org/t/announcing-new-ocean-ddf-survey-strategy-simulation/10162}. The reports on observing strategies can be found at \url{https://survey-strategy.lsst.io/scoc/reports.html}.}(DDF), at the time of writing this publication, suggests one-day cadence observations, rotating two of the \textit{griz} bands per epochs (e.g., $g+r \rightarrow r+i$), during the first year of operation. This, in practice, translates to an effective ${\sim}3$ day cadence for the same filter. In the deep fields, a higher rest-frame cadence will be obtained for high-$z$ \sneia (${\sim}2$ days, comparable to ZTF), probing bluer rest-frame wavelengths and enabling the identification of tens or even hundreds of \eexsne. However, most of the survey will not benefit from this cadence, but will contribute with the discovery and early detection of transients. Nonetheless, the La Silla Schmidt Southern Survey \citep[LS4;][]{Miller+2025} will provide follow up of several LSST transients, for which many will have early-time high-cadence ($1$\,d) observations, enabling future analysis of \eexsne as well. A subset of these will also benefit from early-time spectroscopy, providing more in-depth information of \sneia and their progenitor scenario. However, the challenge of identifying \eexsne early enough to trigger spectroscopic observations at the time of flux excess is difficult to bypass as data during the rise is required.


\section*{Data Availability}

All data (e.g., photometry, light-curve parameters) used as part of this work are already public \citep{Rigault+2025a}. Data as a result of the analysis presented here can be shared upon request to the authors. The analysis codes will be published in the following repository once the article is accepted for publication: \url{https://github.com/temuller/ztf_early_bumps}


\begin{acknowledgements}

We thank the referee for the very constructive report, particularly on the improvements of the statistical methods.
T.E.M.B., K.M., U.B. are funded by Horizon Europe ERC grant no. 101125877.
G.D. acknowledges support from the European Union’s Horizon Europe research and innovation programme under the Marie Skłodowska-Curie grant agreement No 101199369.
L.G. acknowledges financial support from CSIC, MCIN and AEI 10.13039/501100011033 under projects PID2023-151307NB-I00, PIE 20215AT016, and CEX2020-001058-M.
Y.-L.K. was supported by the Lee Wonchul Fellowship, funded through the BK21 Fostering Outstanding Universities for Research (FOUR) Program (grant No. 4120200513819) and the National Research Foundation of Korea to the Center for Galaxy Evolution Research (RS-2022-NR070872, RS-2022-NR070525).
Based on observations obtained with the Samuel Oschin Telescope 48-inch and the 60-inch Telescope at the Palomar Observatory as part of the Zwicky Transient Facility project. ZTF is supported by the National Science Foundation under Grants No. AST-1440341 and AST-2034437 and a collaboration including current partners Caltech, IPAC, the Weizmann Institute of Science, the Oskar Klein Center at Stockholm University, the University of Maryland, Deutsches Elektronen-Synchrotron and Humboldt University, the TANGO Consortium of Taiwan, the University of Wisconsin at Milwaukee, Trinity College Dublin, Lawrence Livermore National Laboratories, IN2P3, University of Warwick, Ruhr University Bochum, Northwestern University and former partners the University of Washington, Los Alamos National Laboratories, and Lawrence Berkeley National Laboratories. Operations are conducted by COO, IPAC, and UW. SED Machine is based upon work supported by the National Science Foundation under Grant No. 1106171.\\

\textit{Software}: 
\texttt{astropy} \citep{astropy1, astropy2, astropy3},
\texttt{corner} \citep{corner}, 
\texttt{emcee} \citep{emcee}, 
\texttt{extinction} \citep{Barbary2016-ext},
\texttt{lmfit} \citep{lmfit},
\texttt{matplotlib} \citep{matplotlib}, 
\texttt{numpy} \citep{numpy},
\texttt{pandas} \citep{pandas},
\texttt{scipy} \citep{scipy},
\texttt{sfdmap}\footnote{\url{https://github.com/kbarbary/sfdmap}}, 
\texttt{ztfdir}\footnote{\url{https://github.com/MickaelRigault/ztfidr}}.

\end{acknowledgements}

\bibliographystyle{aa}
\bibliography{bibliography}

\clearpage
\begin{appendix}

\section{Candidate \eexsne light curves}

In this appendix, we show the candidate \eexsne identified by the methods in Sect.~\ref{subsec:salt2}, \ref{subsec:powerlaw} and ~\ref{subsubsec:ni_dist}.

\begin{figure*}[t]
    \includegraphics[width=\textwidth]{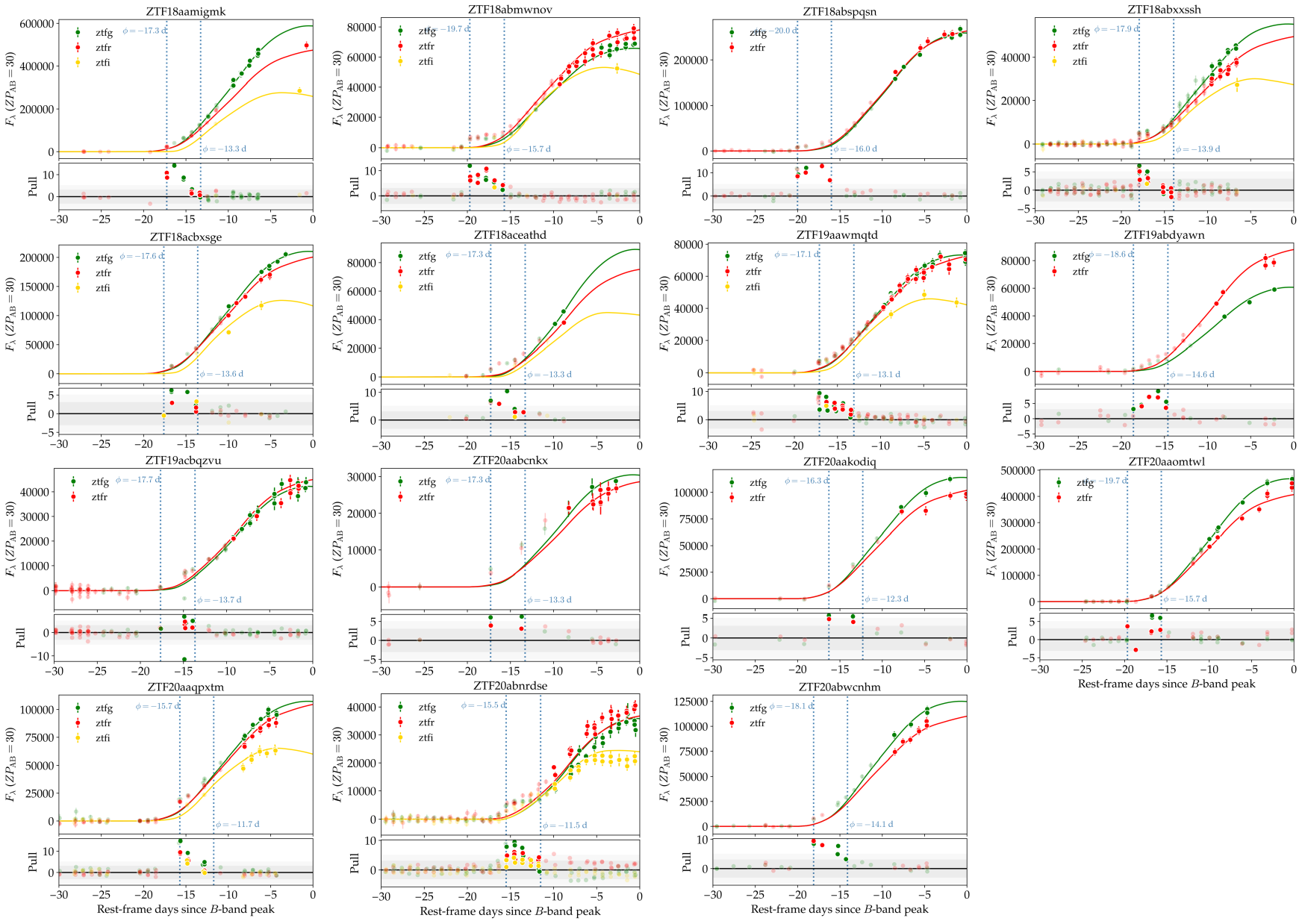}
    \caption{Light curves and SALT2 models for the \eexsn candidates from Sect.~\ref{subsec:salt2}. The rest of the description is as in Fig.~\ref{fig:salt2_example}.}
    \label{fig:salt2_eexsne}
\end{figure*}

\begin{figure*}[t]
    \includegraphics[width=\textwidth]{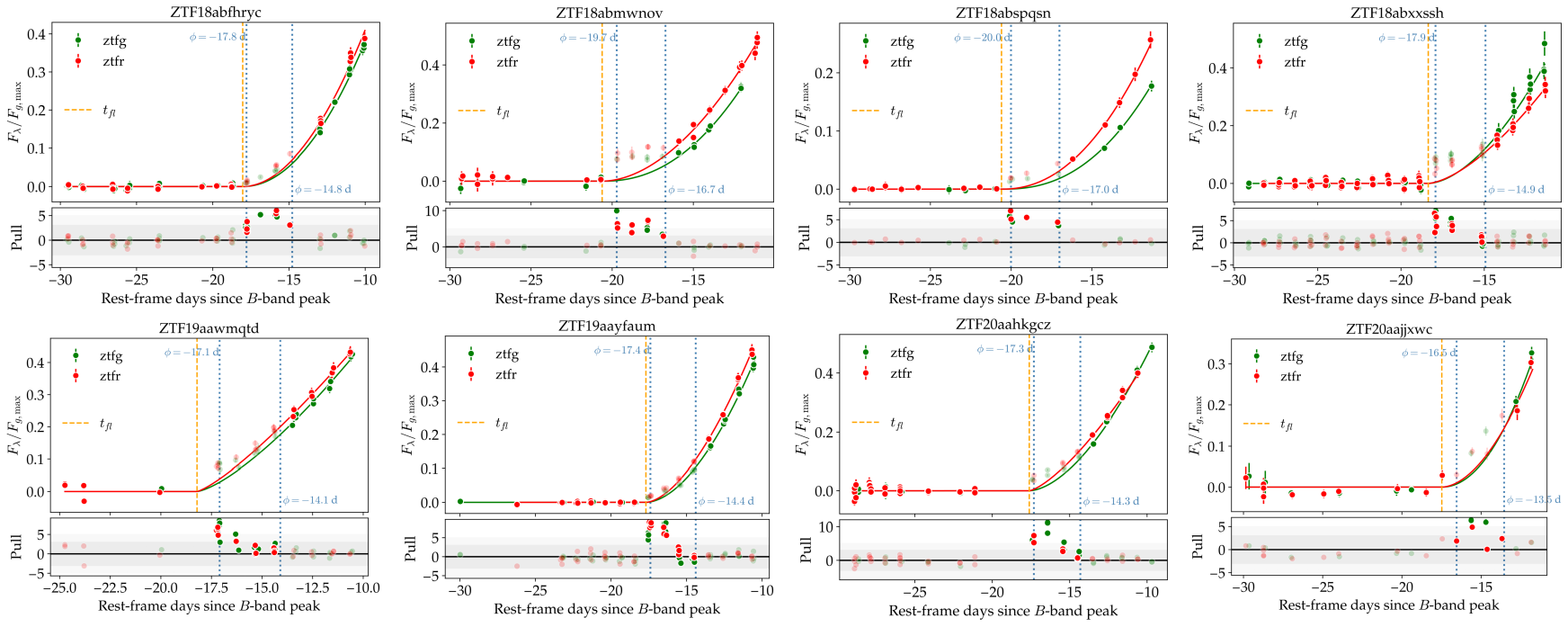}
    \caption{Light curves and power-law models for the \eexsn candidates from Sect.~\ref{subsec:powerlaw}. The rest of the description is as in Fig.~\ref{fig:powerlaw_example}.}
    \label{fig:powerlaw_eexsne}
\end{figure*}

\begin{figure*}[t]
    \includegraphics[width=\textwidth]{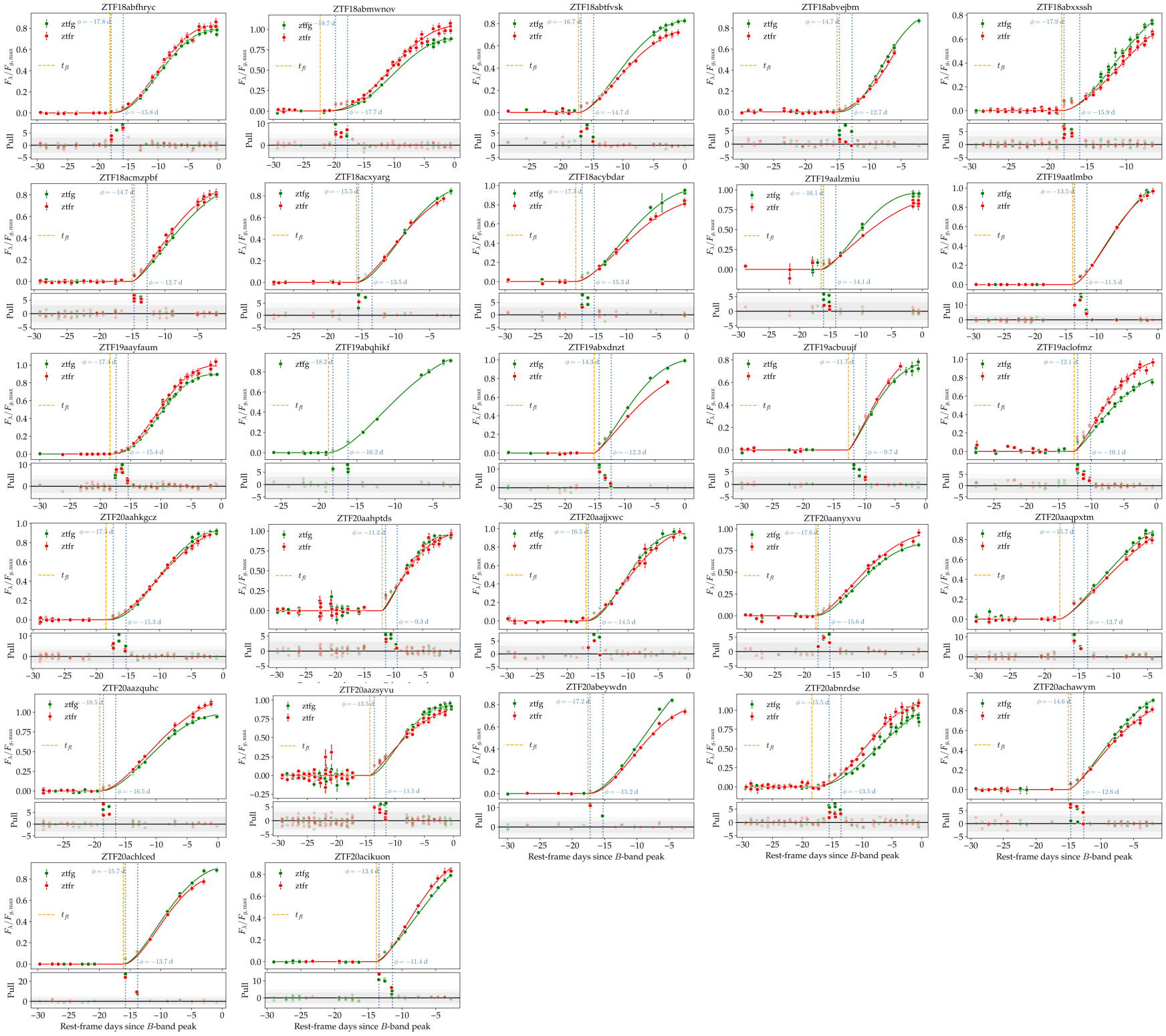}
    \caption{Light curves and modified power-law models for the \eexsn candidates from Sect.~\ref{subsec:powerlaw}. The rest of the description is as in Fig.~\ref{fig:powerlaw_example}.}
    \label{fig:modified_powerlaw_eexsne}
\end{figure*}

\begin{figure*}[t]
    \includegraphics[width=\textwidth]{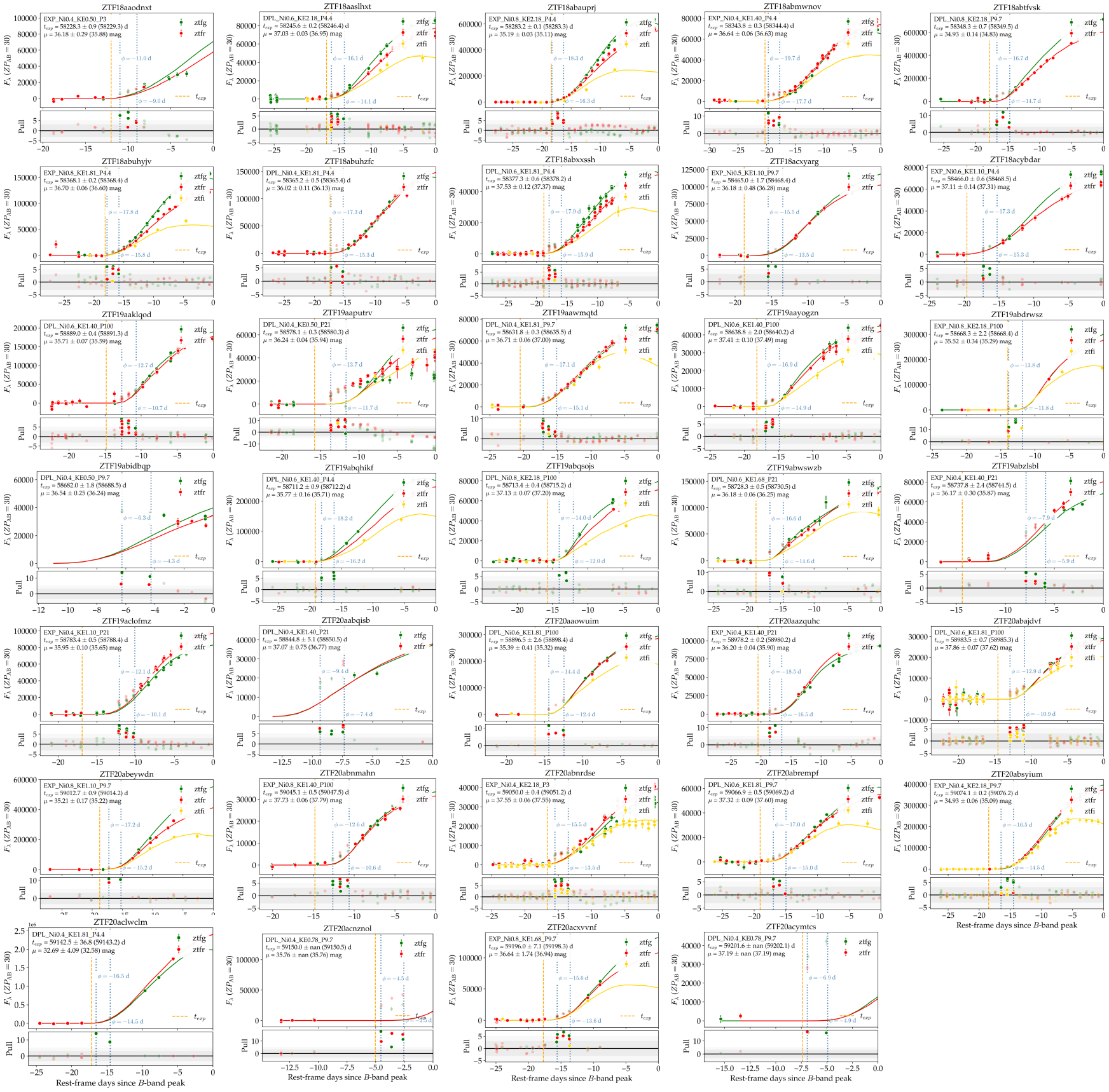}
    \caption{Light curves and \nickel-distribution models for the \eexsn candidates from Sect.~\ref{subsubsec:ni_dist}. The best-fit model and its parameters, including uncertainties, are shown, together with the initial guess in parenthesis. The rest of the description is as in Fig.~\ref{fig:turtls_example}.}
    \label{fig:nickel_dist_eexsne}
\end{figure*}

\section{Comparison to the ZTF 2018 sample}

In Table~\ref{tab:2018_sne}, we summarise the \eexsn candidates from previous works, using the ZTF 2018 sample, compared to those identified in this work.

\begin{table*}[h]
\caption{\sneia discovered by ZTF in 2018 considered to be \eexsne by previous works.}
\centering
\begin{threeparttable}

\begin{tabular}{lcccccc}
\hline
ZTF name & IAU name & Sect.\ref{subsec:models_without_excess}$^\star$ & Sect.\ref{subsec:models_with_excess}$^\diamond$ & Passing Cuts$^\clubsuit$ & Subtype$^\flat$ & Reference$^\dagger$ \\
\hline
ZTF18aavrwhu & 2018bxo &  & DD & yes & norm & 5 \\
ZTF18abxxssh & 2018gvj & S,PL,MPL,ND & DD & yes & norm & 1,2,4,5 \\
ZTF18aapqwyv & 2018bhc &  &  & fitquality & norm & 2 \\
ZTF18abcflnz & 2018cuw &  & CI & yes & norm & 2,5 \\
ZTF18abckujq & 2018cvf &  &  & yes & norm & 2 \\
ZTF18abcrxoj & 2018cvw &  &  & yes & norm & 2 \\
ZTF18abgxvra & 2018efb &  &  & yes & norm & 2 \\
ZTF18aaqqoqs & 2018cbh &  & DD & yes & 91T & 4 \\
ZTF18aayjvve & 2018cny &  &  & yes & norm & 4 \\
ZTF18abdfazk & 2018dbe &  &  & yes & norm & 4,5 \\
ZTF18abdfwur & 2018eps &  & CI & fitquality & unknown & 4 \\
ZTF18abpamut & 2018fqe &  &  & fitquality & 91T & 4 \\
ZTF18aaqcozd & 2018bjc &  &  & yes & norm & 5 \\
ZTF18aawjywv & 2018ccj &  &  & not in DR2 & -- & 5 \\
ZTF18aaxsioa & 2018cfa &  &  & yes & norm & 5 \\
ZTF18aazsabq & 2018crn &  & CI & yes & norm & 5 \\
ZTF18abfhryc & 2018dhw & PL,MPL & DD,CI & yes & norm & 5 \\
ZTF18abimsyv & 2018eni &  & DD & yes & norm & 5 \\
ZTF18abssuxz & 2018gfe &  &  & yes & norm & 5 \\
\hline
\end{tabular}

\begin{tablenotes}
 \item \textbf{Notes.} 
 $^\star$ S: SALT2, PL: Power Law, MPL: Modified Power Law, ND: \nickel Distribution.\\
 $^\diamond$ DD: Double Detonation., CI: Companion Interaction.\\
 $^\clubsuit$ The cut the SN fails to pass is shown, if any.\\
 $^\flat$ According to ZTF DR2.\\
 $^\dagger$1: \cite{Yao+2019}, 2: \cite{Bulla+2020}, 3: \citetalias{Miller+2020}, 4 \citetalias{Deckers+2022}, and 5: \cite{Burke+2022}.
\end{tablenotes}

\end{threeparttable}
\label{tab:2018_sne}

\end{table*}

\section{Rise-time observations}

\begin{figure}[!ht]
    \includegraphics[width=\columnwidth]{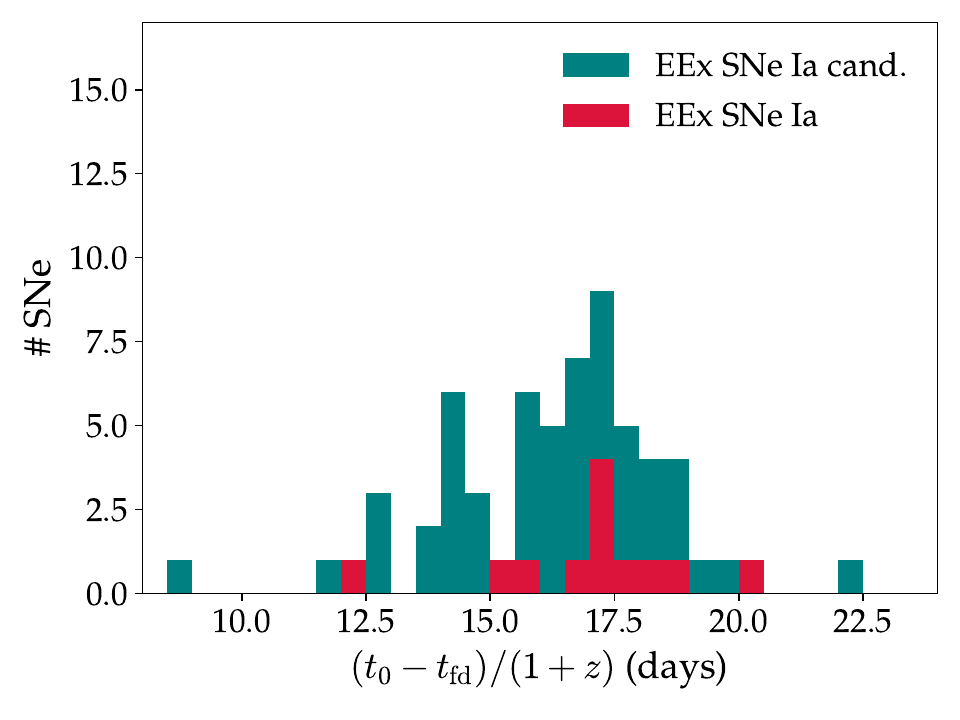}
    \includegraphics[width=\columnwidth]{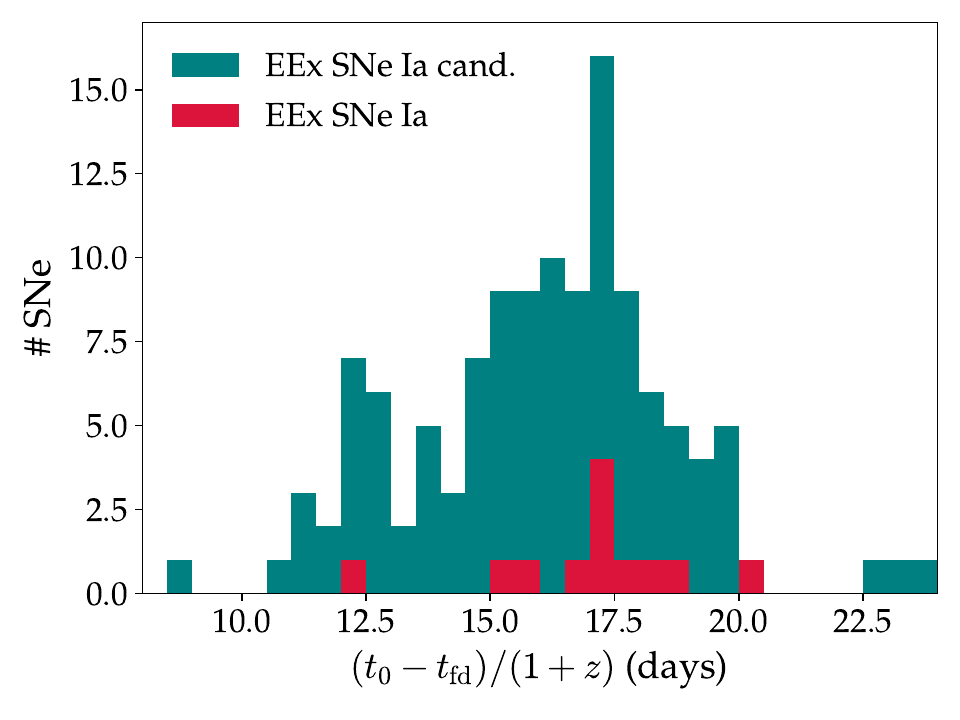}
    \caption{Similar to Fig.~\ref{fig:rise_from_tfd} but for \eexsn candidates from Sect.~\ref{subsec:models_without_excess} and \ref{subsec:models_with_excess}. The distribution for the candidates identified by the DD method are shown in the top panel, while those from the CI method are shown in the bottom panel. Note the same range used in both panels.}
    \label{fig:bumpy_rise_from_tfd}
\end{figure}

The pseudo-rise time distribution of \eexsn candidates is shown in Fig.~\ref{fig:bumpy_rise_from_tfd}.

\begin{figure*}[t]
    \includegraphics[width=\columnwidth]{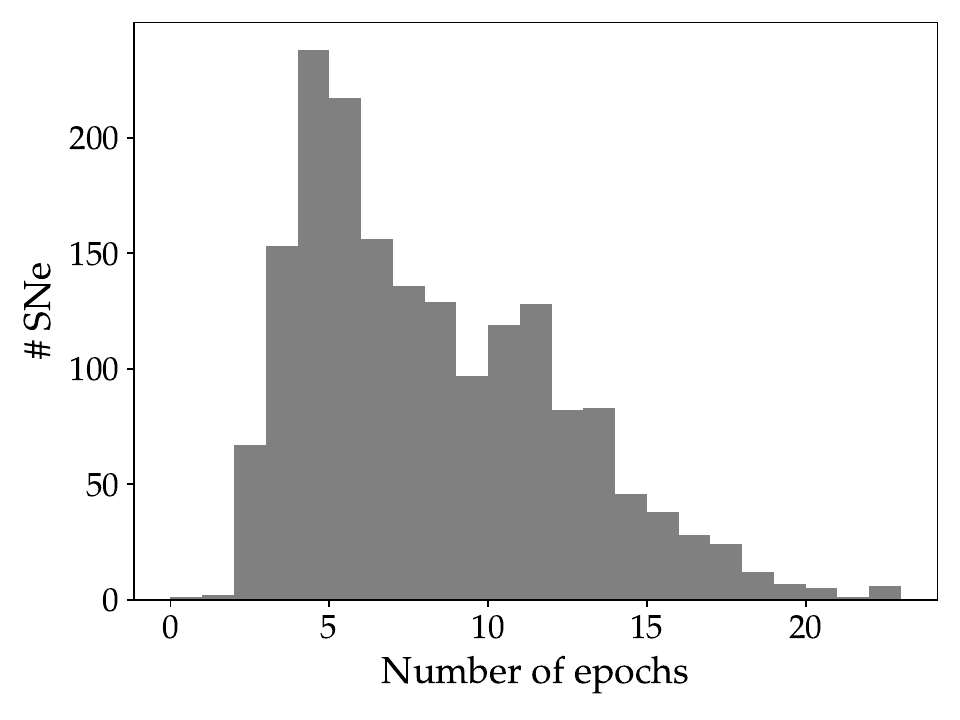}
    \includegraphics[width=\columnwidth]{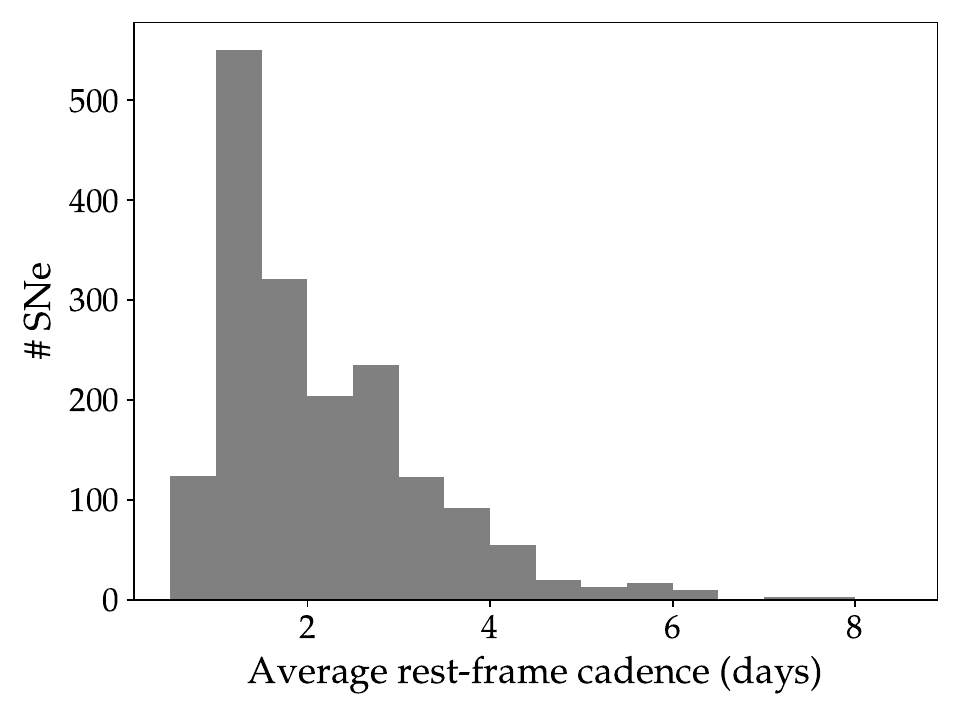}
    \caption{Same as Fig.~\ref{fig:rise_and_cadence} but for \sneia not within the best \eexsn candidates. A bimodal distribution is seen for the number of epochs (\textit{left} panel), which is caused by the observing strategy of ZTF (\smith). The average cadence distribution (\textit{right} panel) peaks at $\mathord{\sim}1.5$\,d and rapidly decreases, with a median of $\mathord{\sim}2$\,d.}
    \label{fig:rest_rise_and_cadence}
\end{figure*}

The distribution of number of epochs during rise time of non-\eexsne is shown in Fig.~\ref{fig:rest_rise_and_cadence}. This is described by a bimodal distribution, with a narrow population that peaks around 5 epochs, and a wider population that peaks around 10 epochs. The average cadence distribution peaks at $\mathord{\sim}1.5$\,d and rapidly decreases, with a median of $\mathord{\sim}2$\,d.

\section{Early colour evolution}

\begin{figure}[!ht]
    \includegraphics[width=\columnwidth]{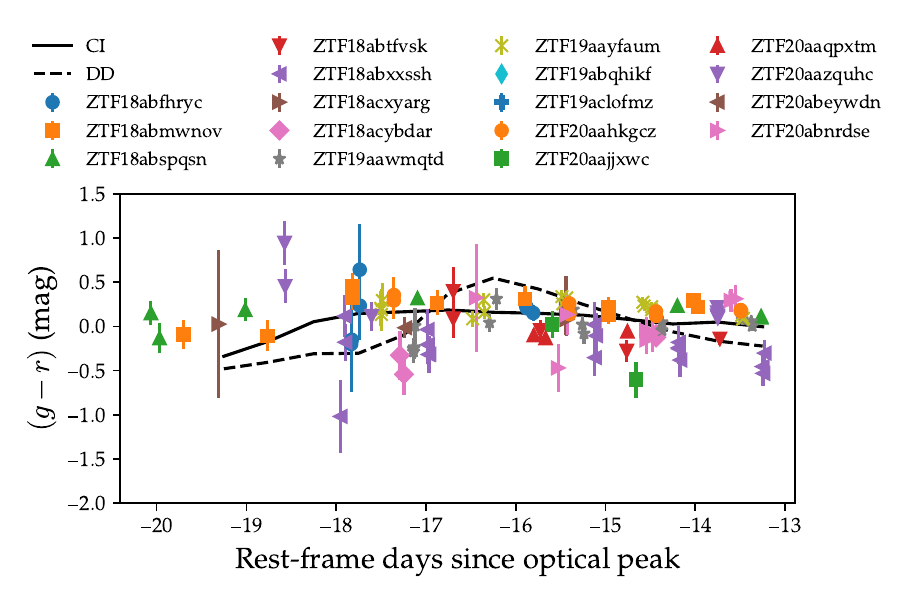}
    \caption{Early $(g-r)$ colour evolution of the best \eexsn candidates. DD (\texttt{WD1.00\_He0.07\_BF0.20\_BP56ni}) and CI (\texttt{Ni0.5\_KE1.10\_s9.7\_a2.00e+12\_Theta135.0}) models are shown for comparison, assuming $t_0-t_{\rm{fl}}=20$ rest-frame days.}
    \label{fig:early_colours}
\end{figure}

The early $(g-r)$ colour evolution of the best \eexsn candidates is shown in Fig.~\ref{fig:early_colours}.

\section{Parameter distributions}

\begin{figure*}[t]
    \includegraphics[width=0.5\columnwidth]{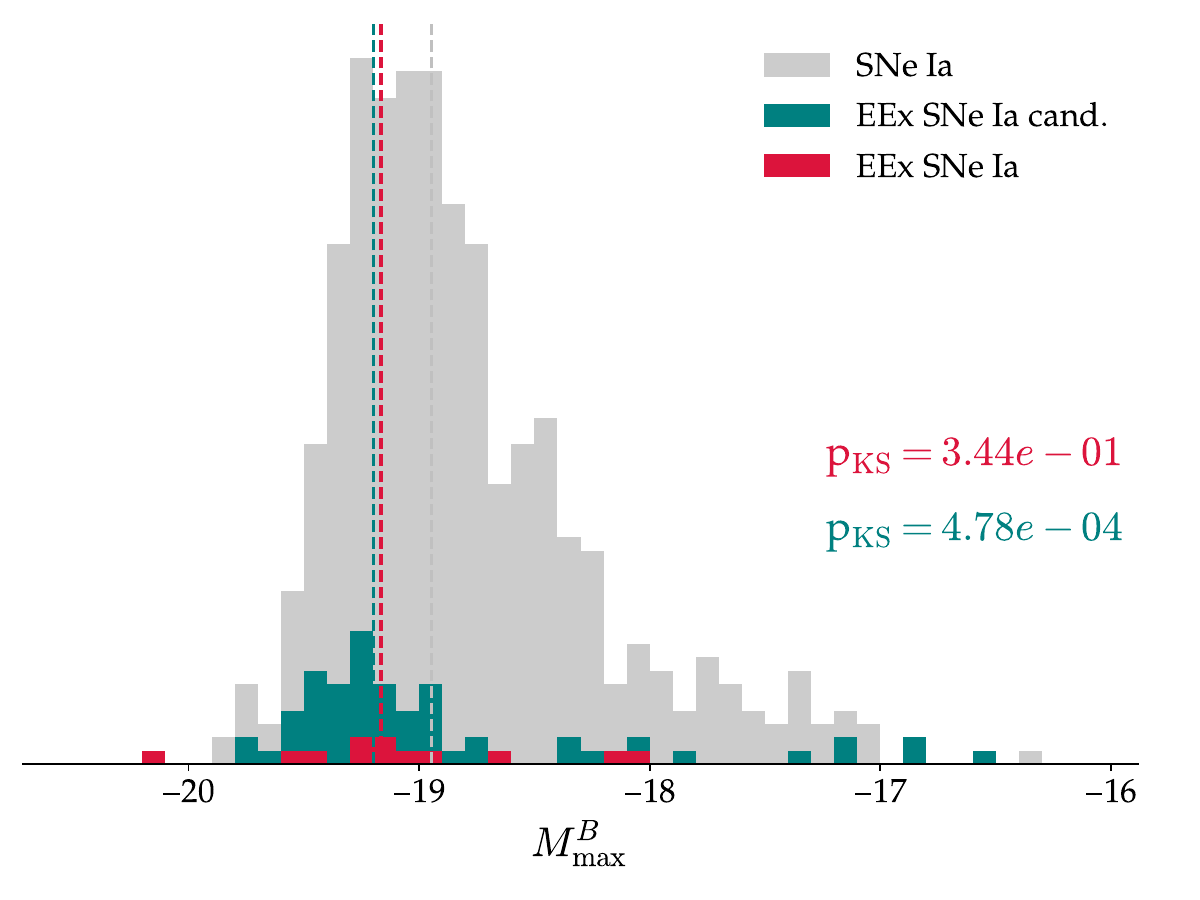}
    \includegraphics[width=0.5\columnwidth]{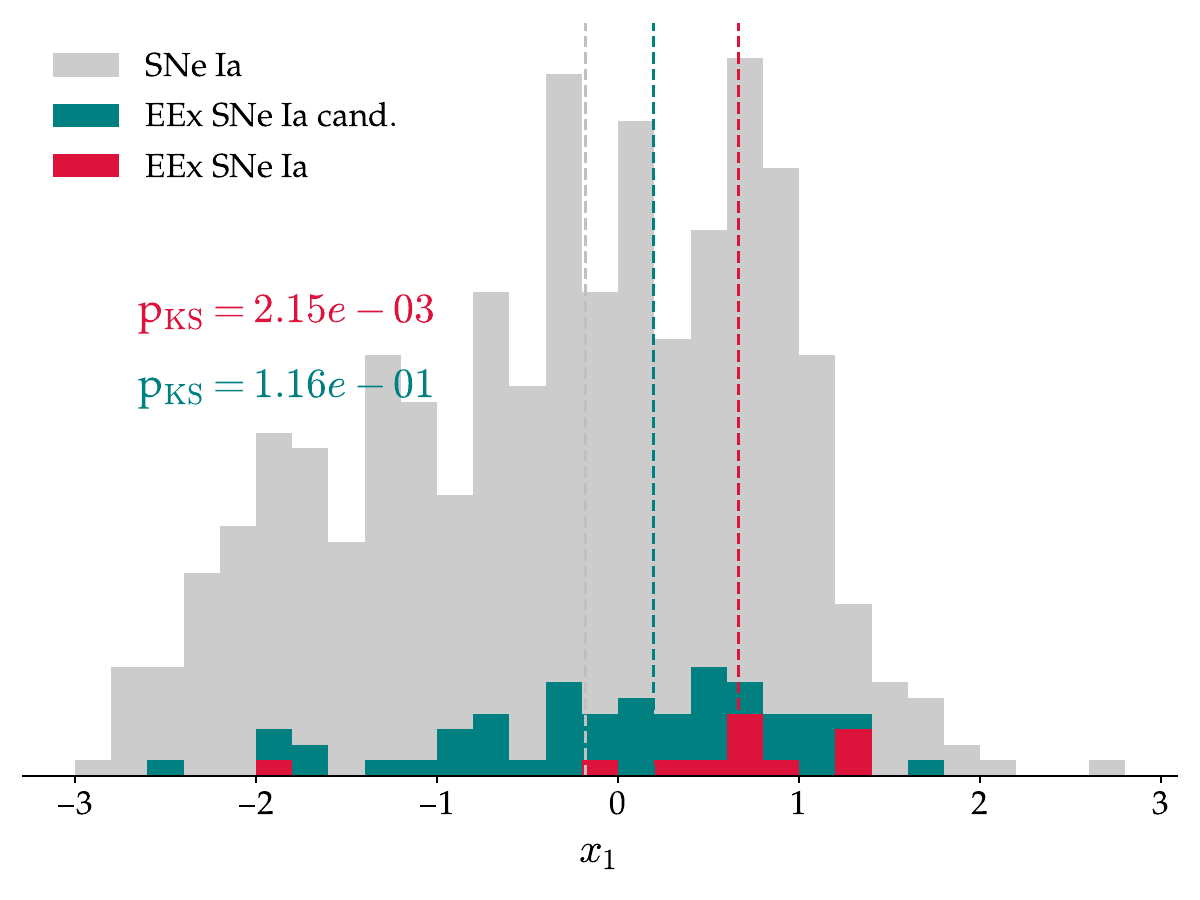}
    \includegraphics[width=0.5\columnwidth]{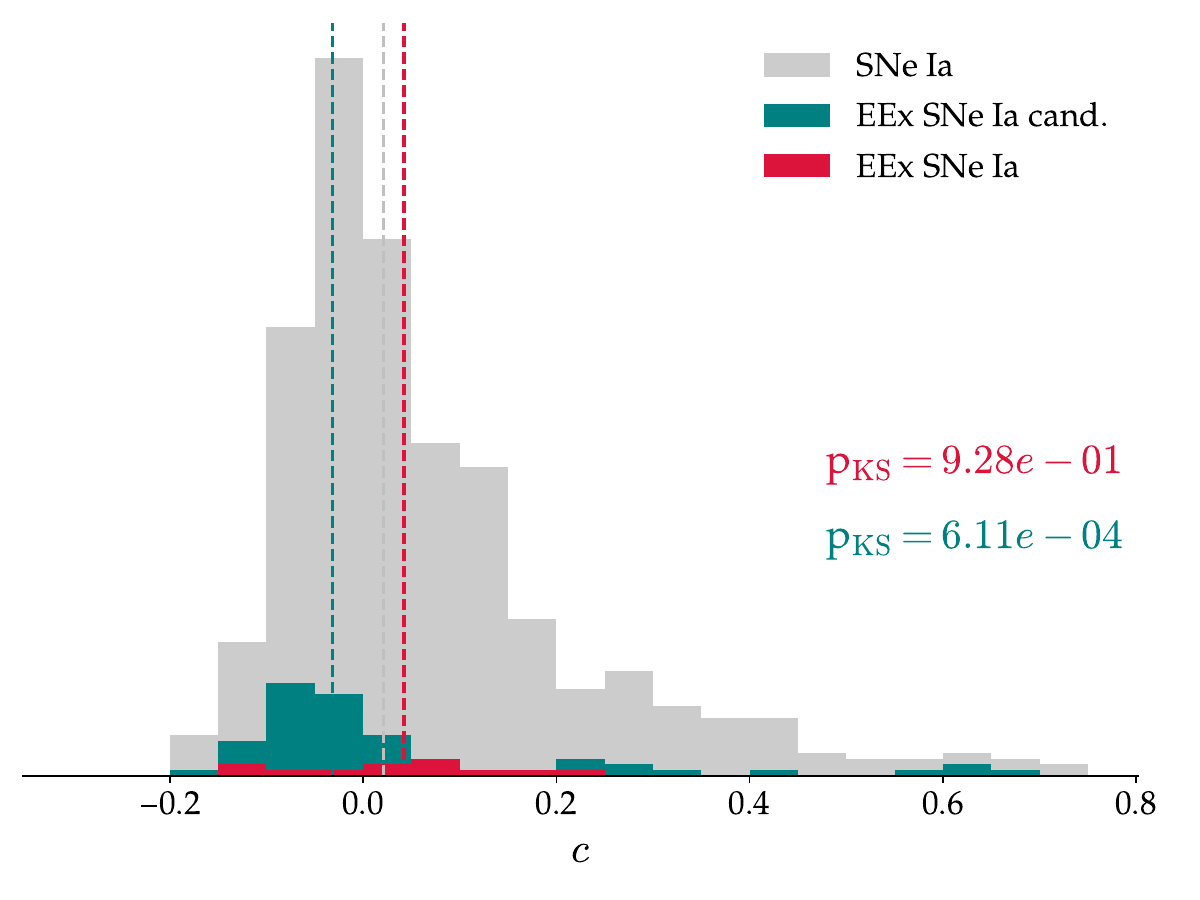}
    \includegraphics[width=0.5\columnwidth]{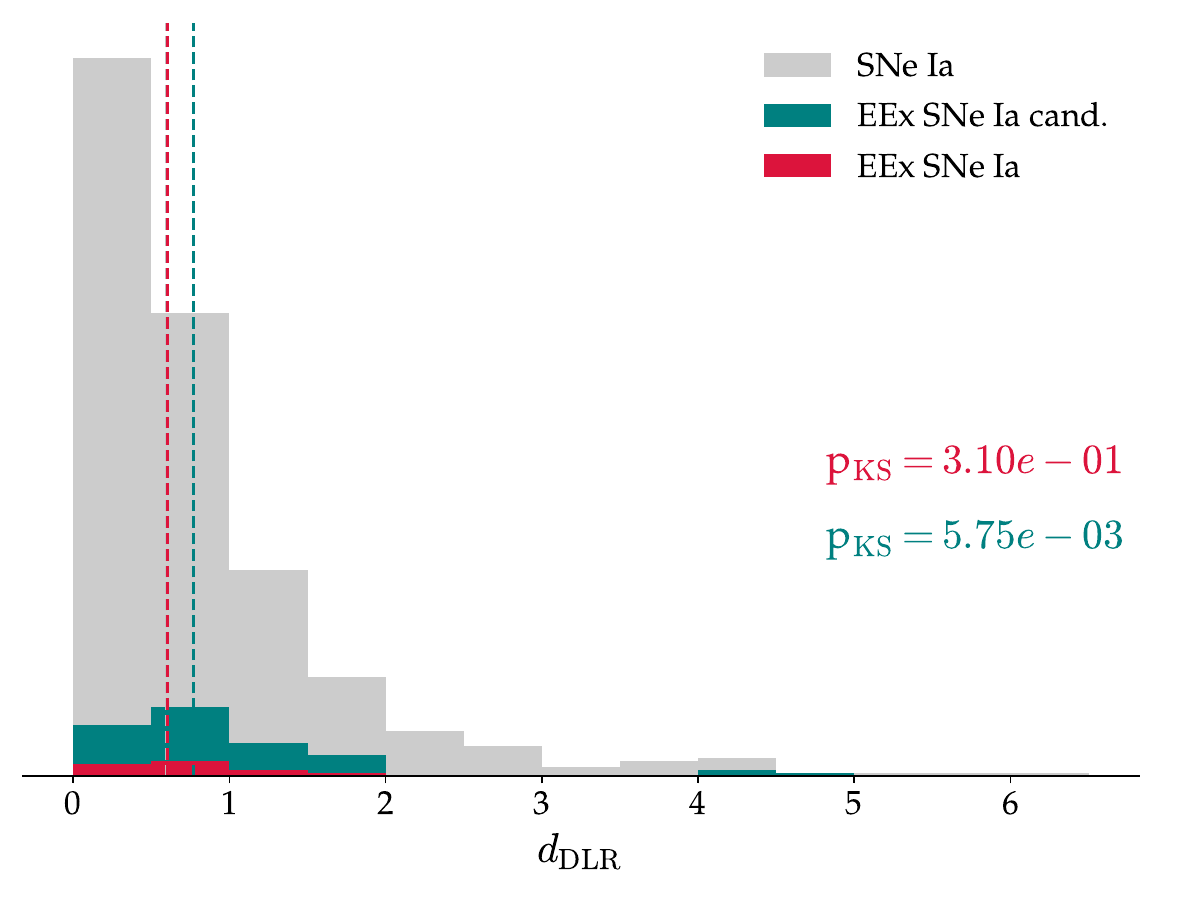}
    \includegraphics[width=0.5\columnwidth]{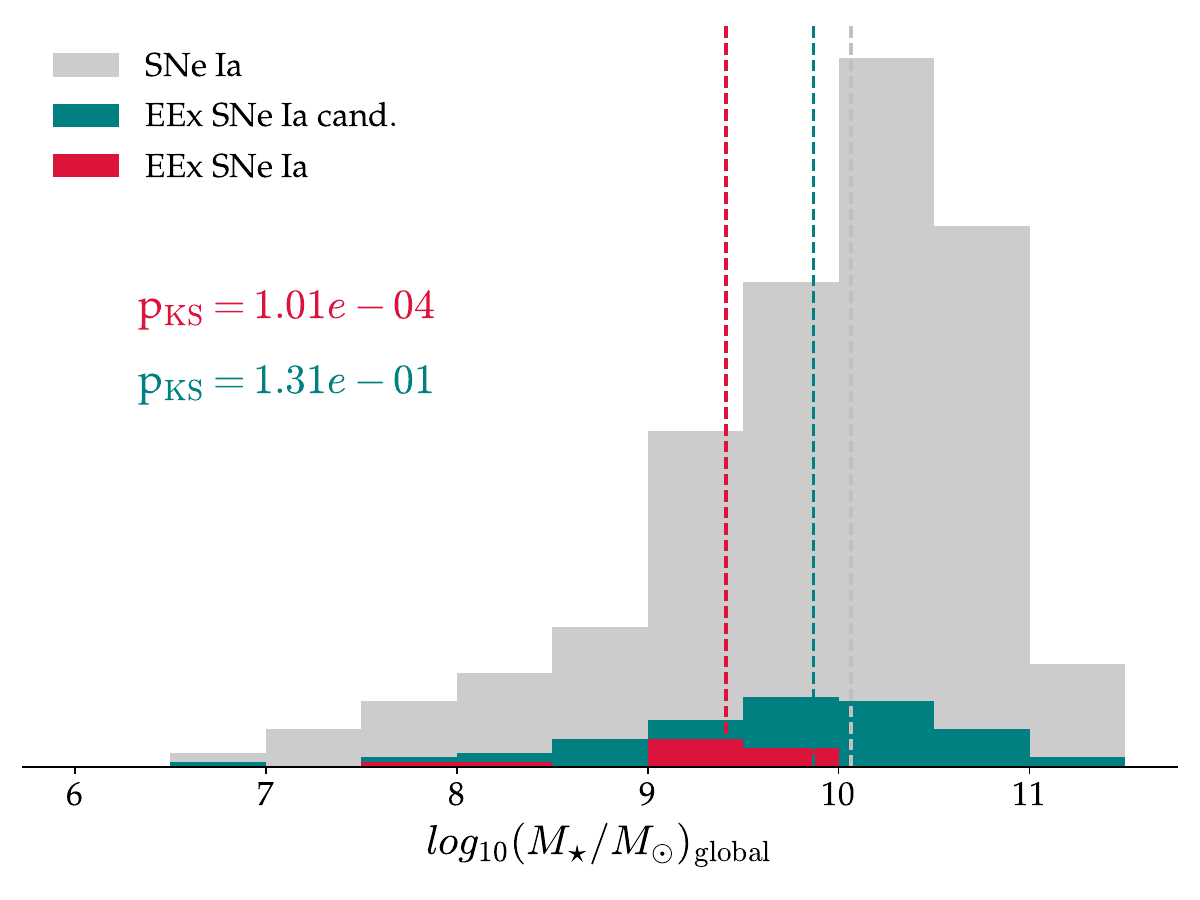}
    \includegraphics[width=0.5\columnwidth]{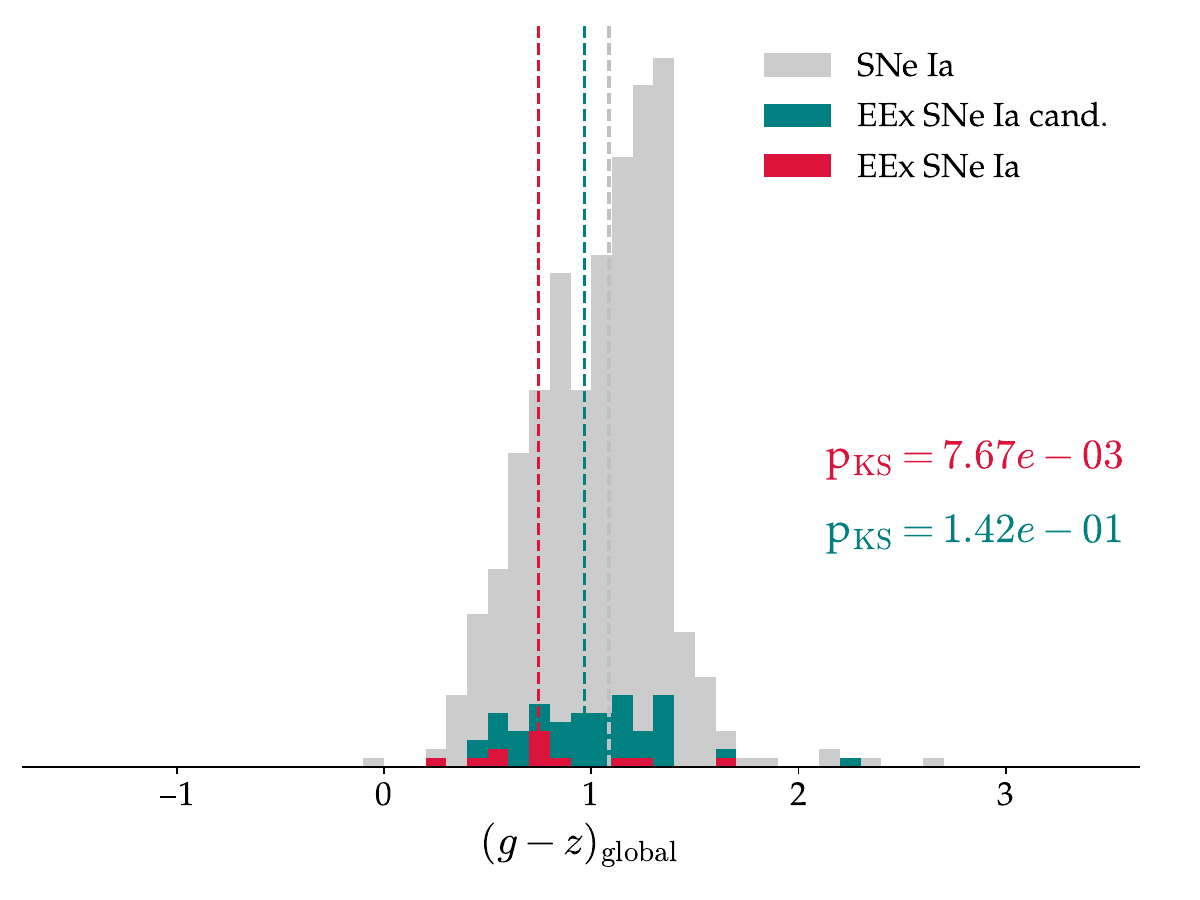}
    \includegraphics[width=0.5\columnwidth]{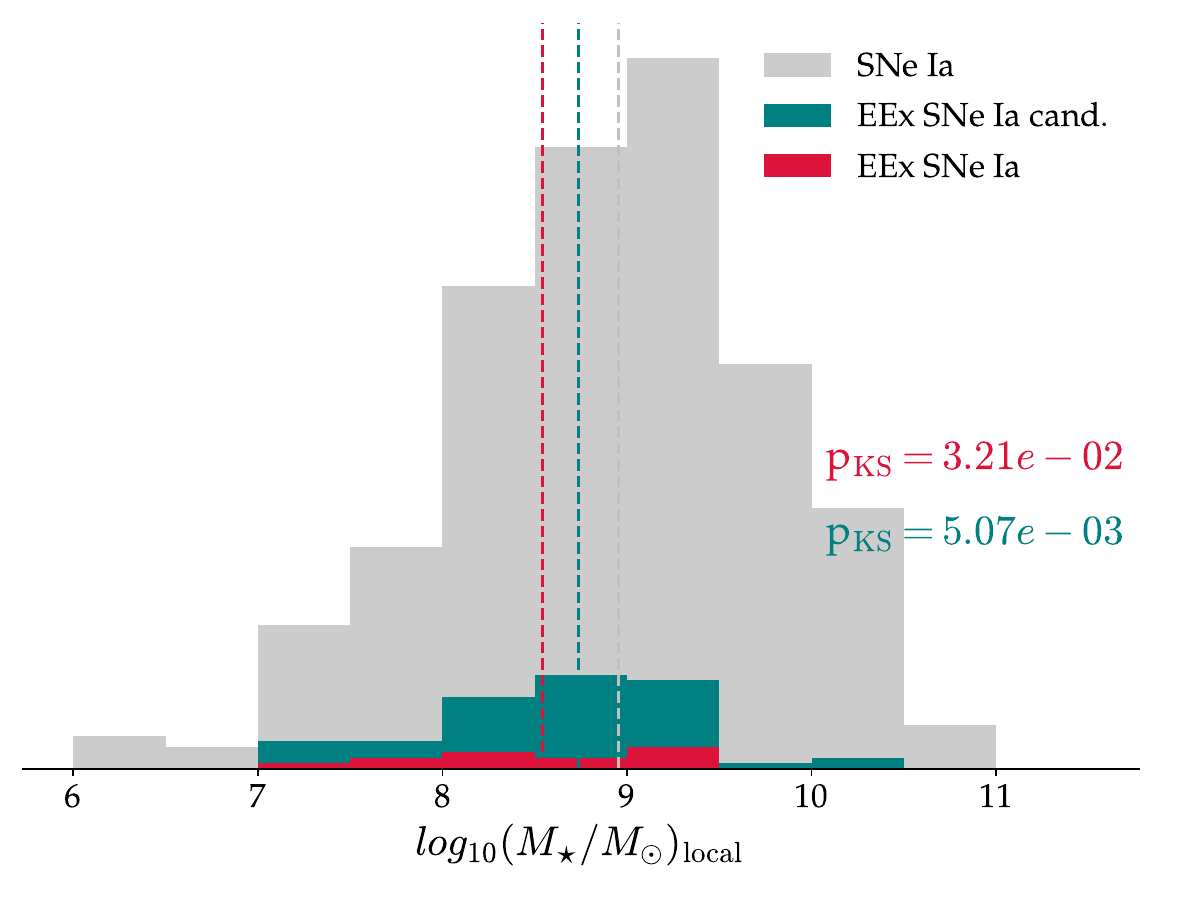}
    \includegraphics[width=0.5\columnwidth]{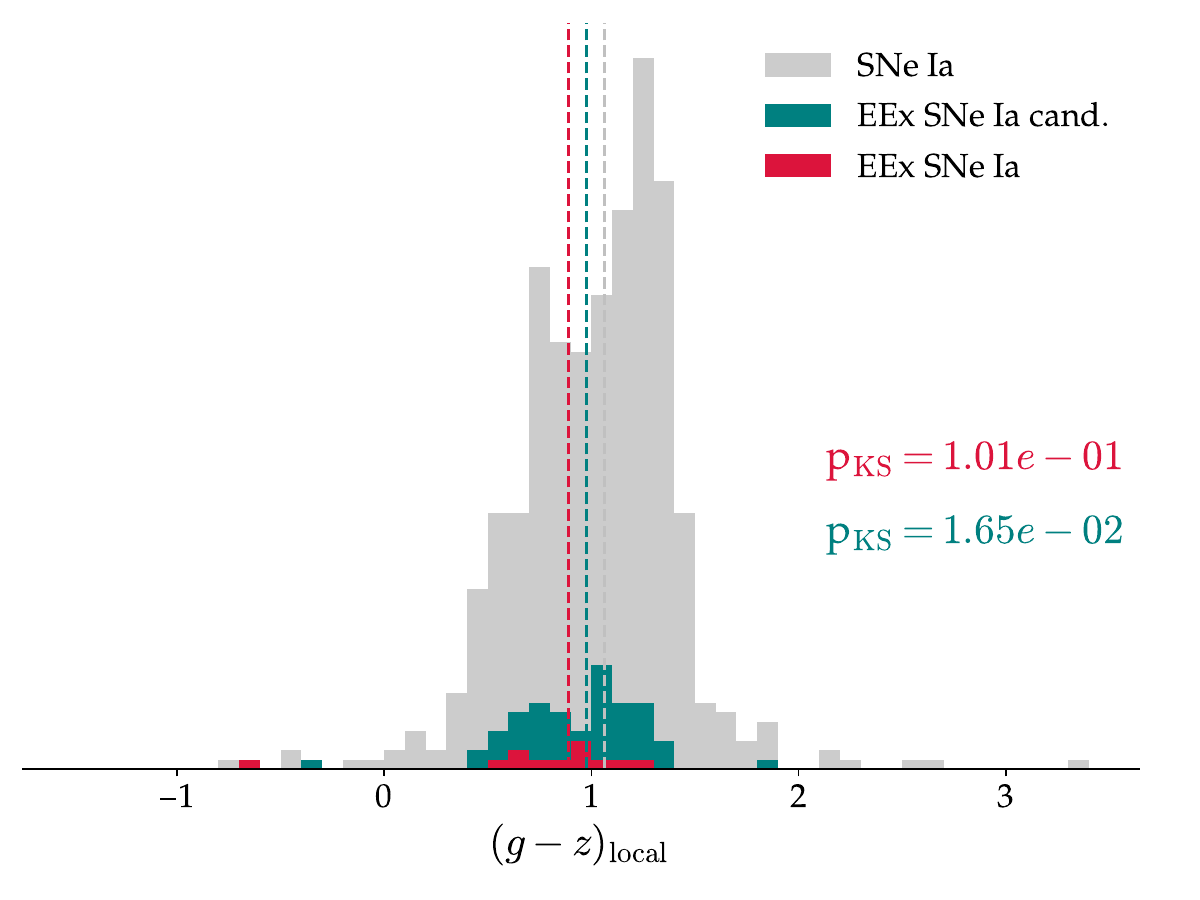}
    \caption{Distributions of light-curve parameters and host-galaxy properties of the SNe Ia in this work: non-\eexsne (grey), best (crimson) and all \eexsn candidates for the DD method (teal). See Sect.~\ref{subsec:parameters_distributions} for more information. The p-value of the KS test (p$_{\rm KS}$), comparing with the non-\eexsne, is shown for each distribution.}
    \label{fig:distributions_DD}
\end{figure*}

\begin{figure*}[t]
    \includegraphics[width=0.5\columnwidth]{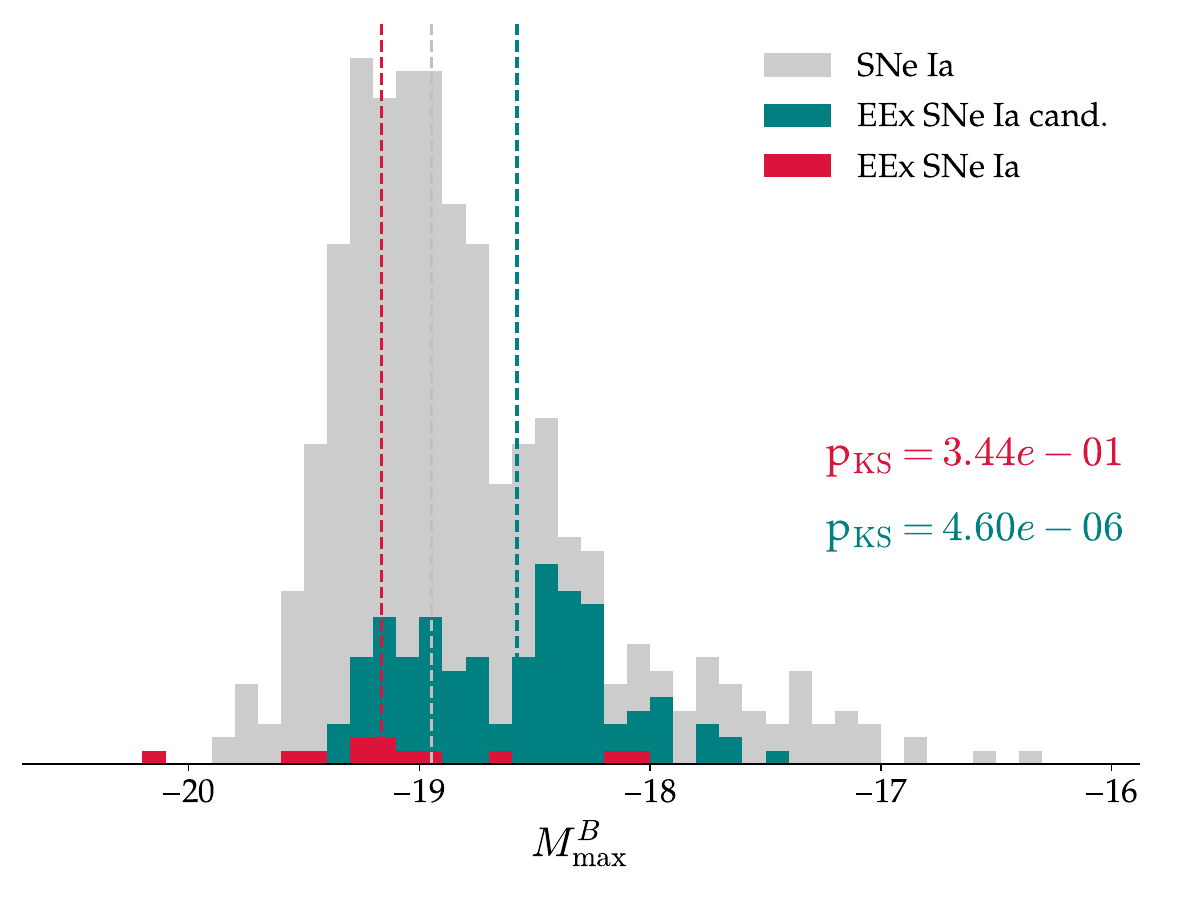}
    \includegraphics[width=0.5\columnwidth]{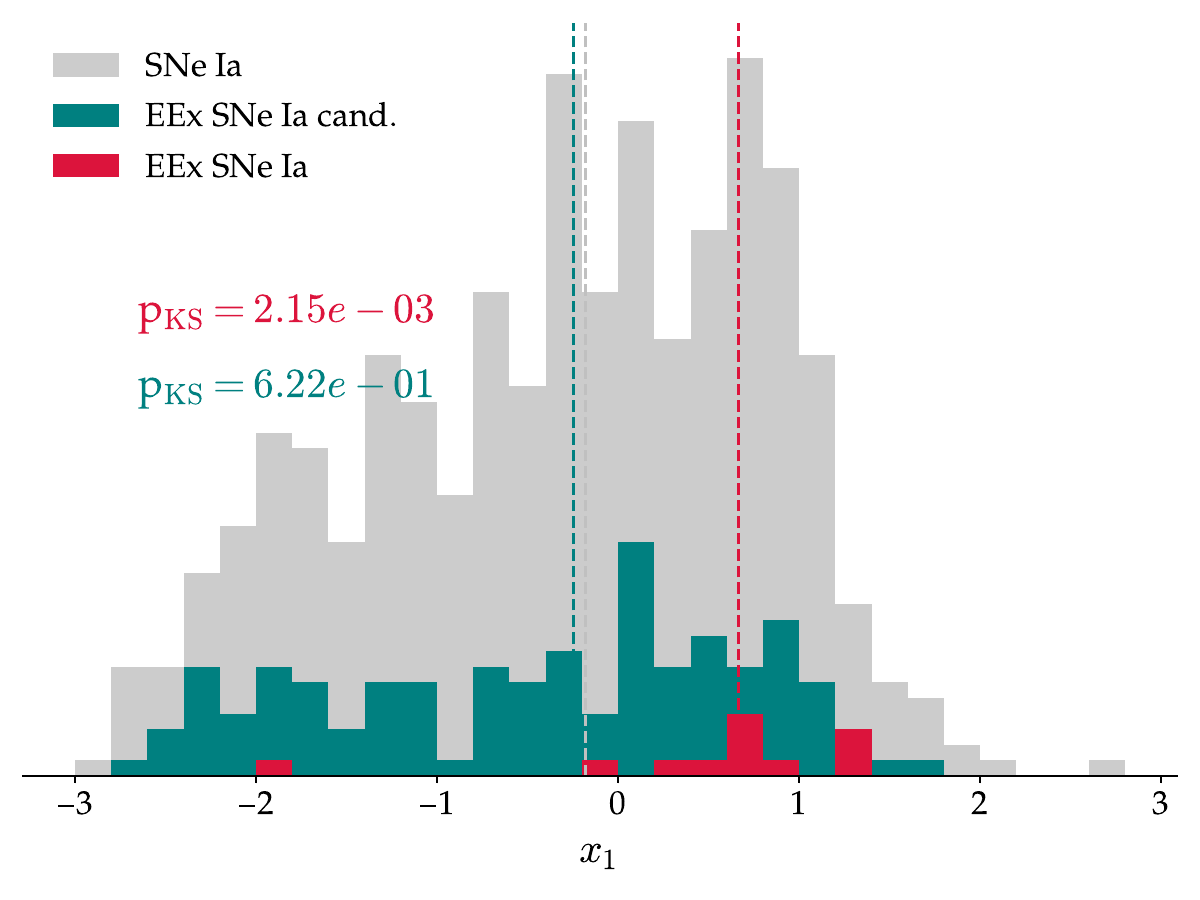}
    \includegraphics[width=0.5\columnwidth]{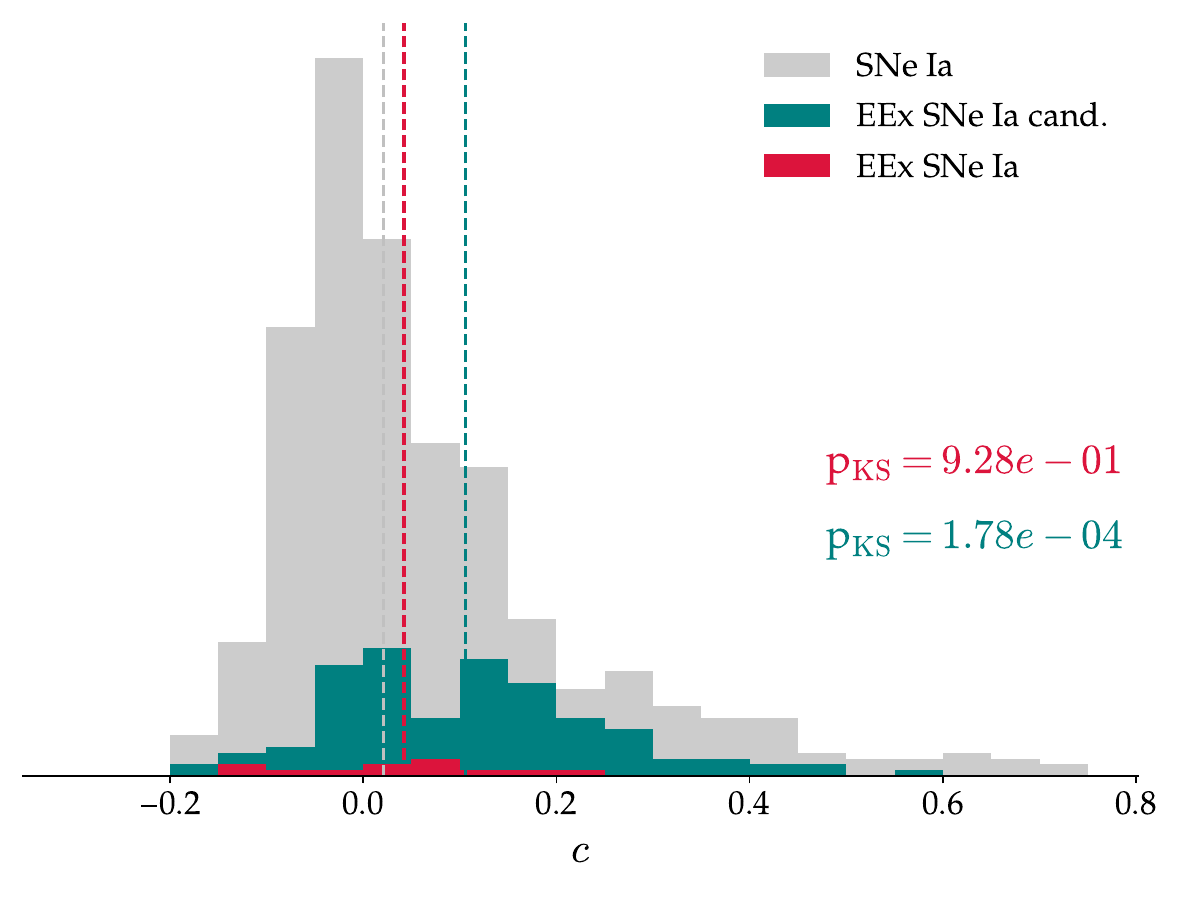}
    \includegraphics[width=0.5\columnwidth]{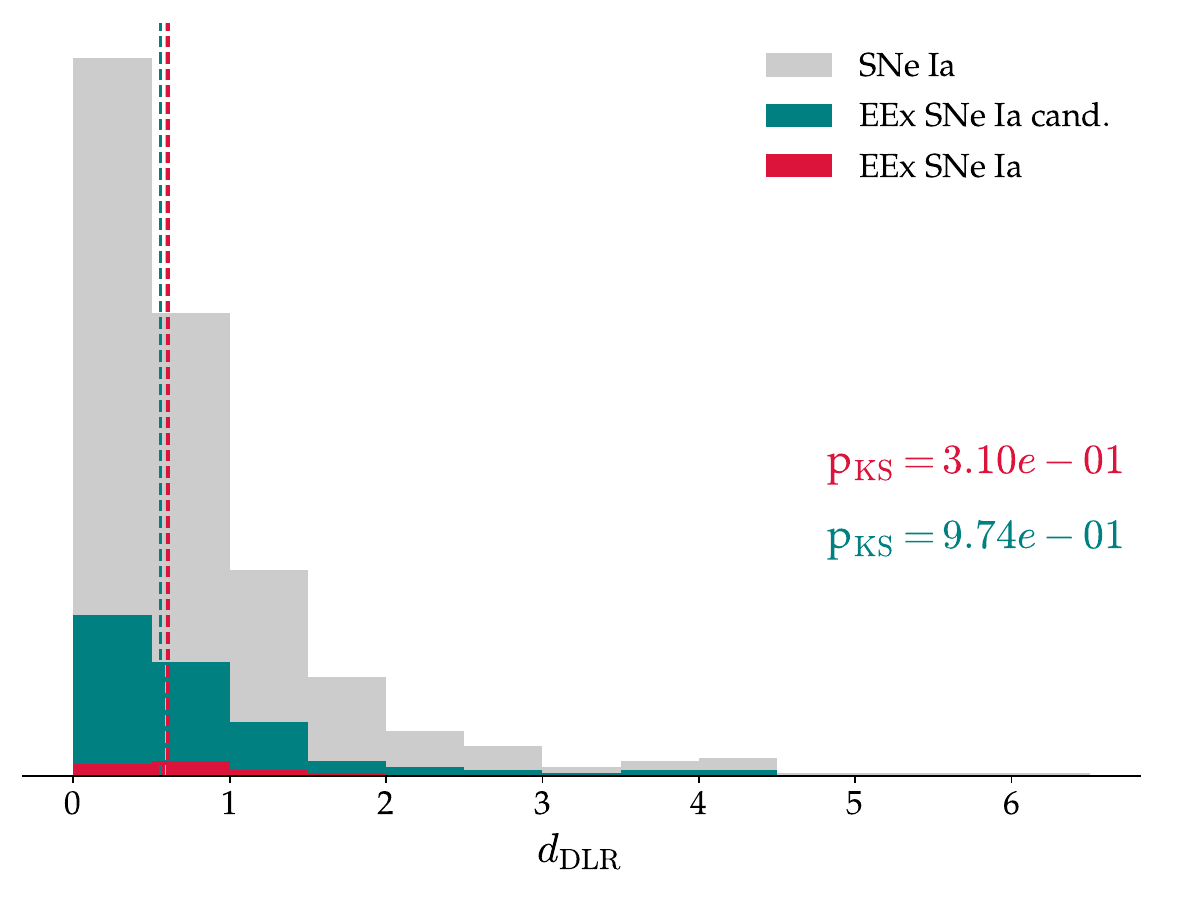}
    \includegraphics[width=0.5\columnwidth]{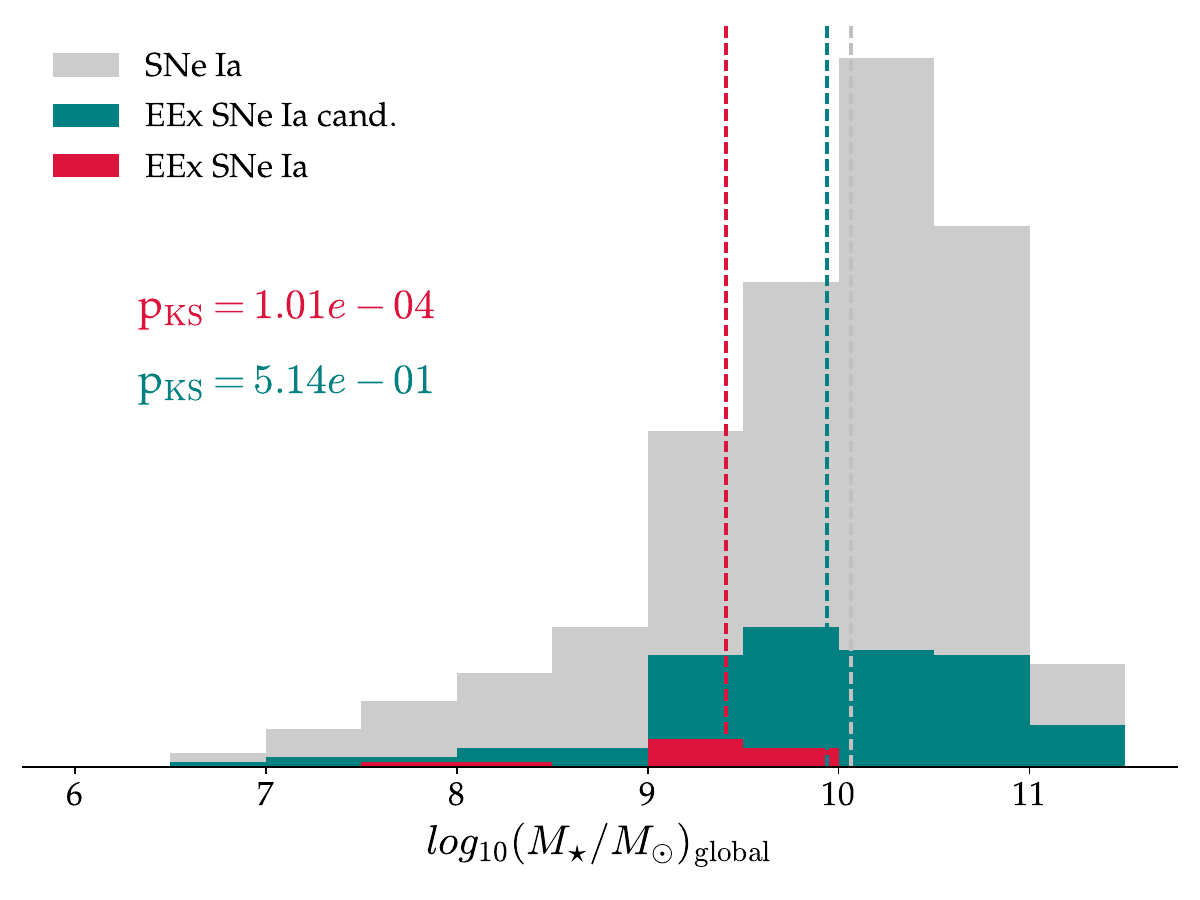}
    \includegraphics[width=0.5\columnwidth]{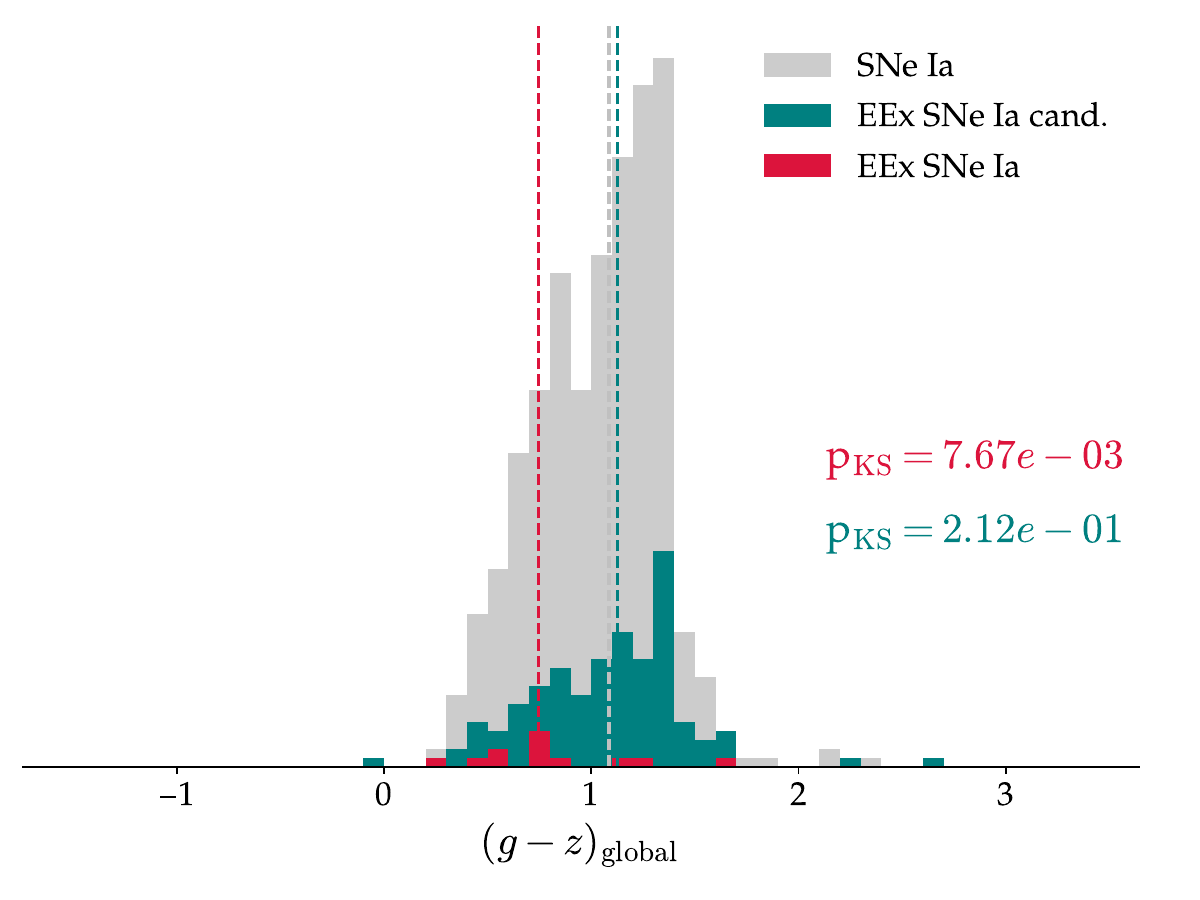}
    \includegraphics[width=0.5\columnwidth]{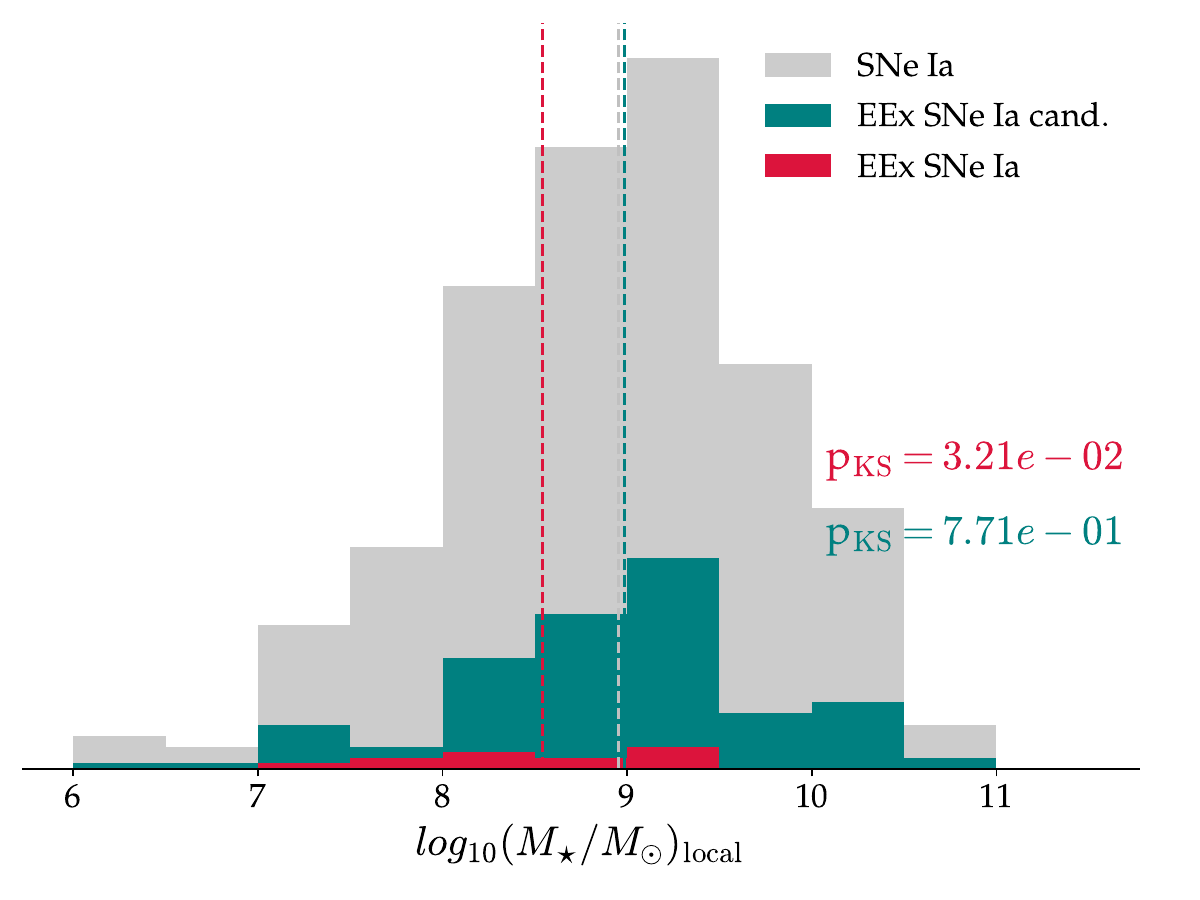}
    \includegraphics[width=0.5\columnwidth]{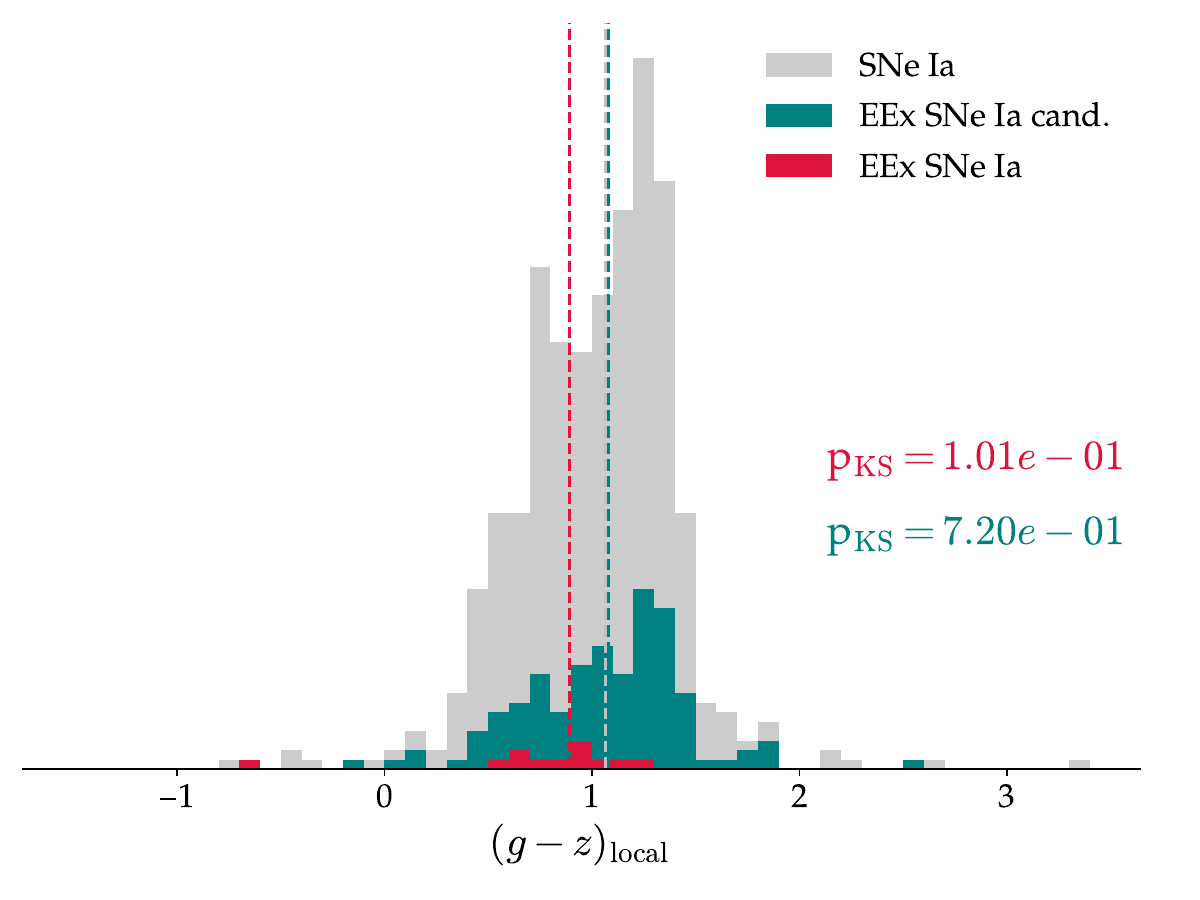}
    \caption{Same as Fig.~\ref{fig:distributions_DD} but for the CI method.}
    \label{fig:distributions_CI}
\end{figure*}

The distributions of the light-curve parameters and host-galaxy properties of the SNe used in this work are presented in Fig.~\ref{fig:distributions_DD} and \ref{fig:distributions_CI}.

\section{Double-detonation and companion-interaction parameters for non-\eexsne}

\begin{figure}[!ht]
    \includegraphics[width=\columnwidth]{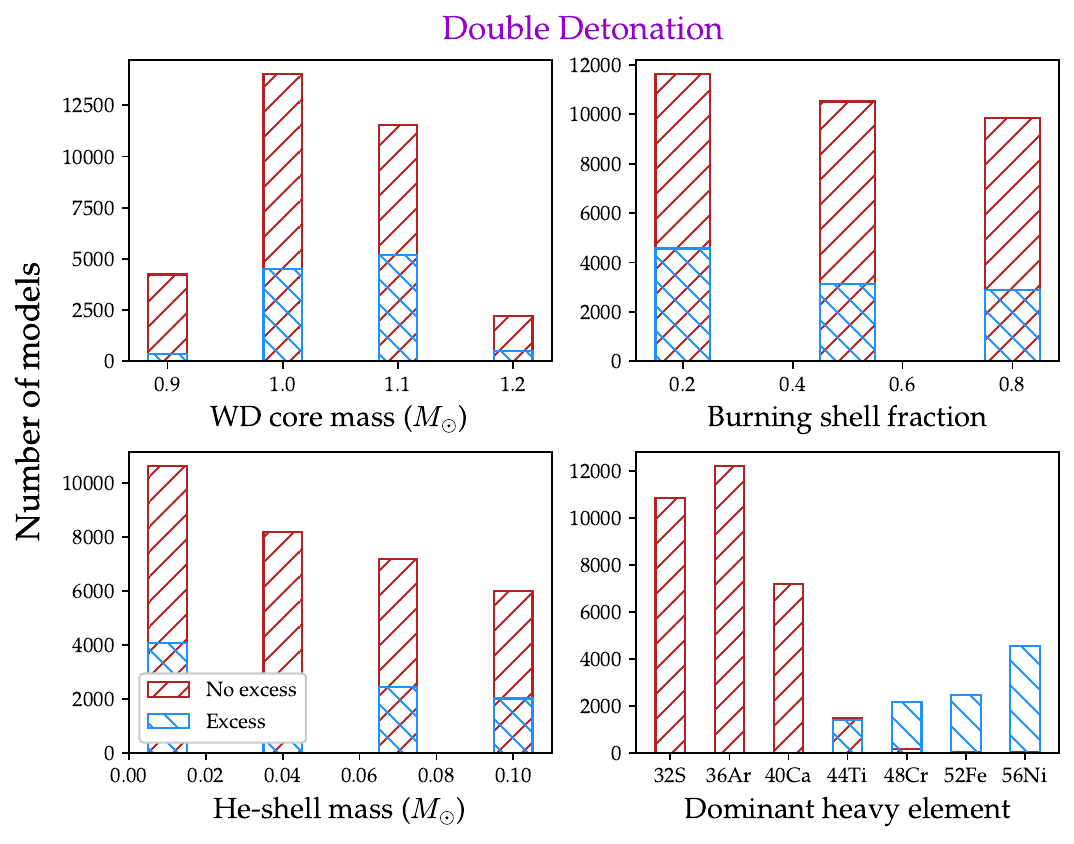}
    \caption{Similar to Fig.~\ref{fig:parameter_distributions_DoubleDetonation} but for non-\eexsne.}
    \label{fig:parameter_distributions_nonEExSNe_DoubleDetonation}
\end{figure}

\begin{figure}[!ht]
    \includegraphics[width=\columnwidth]{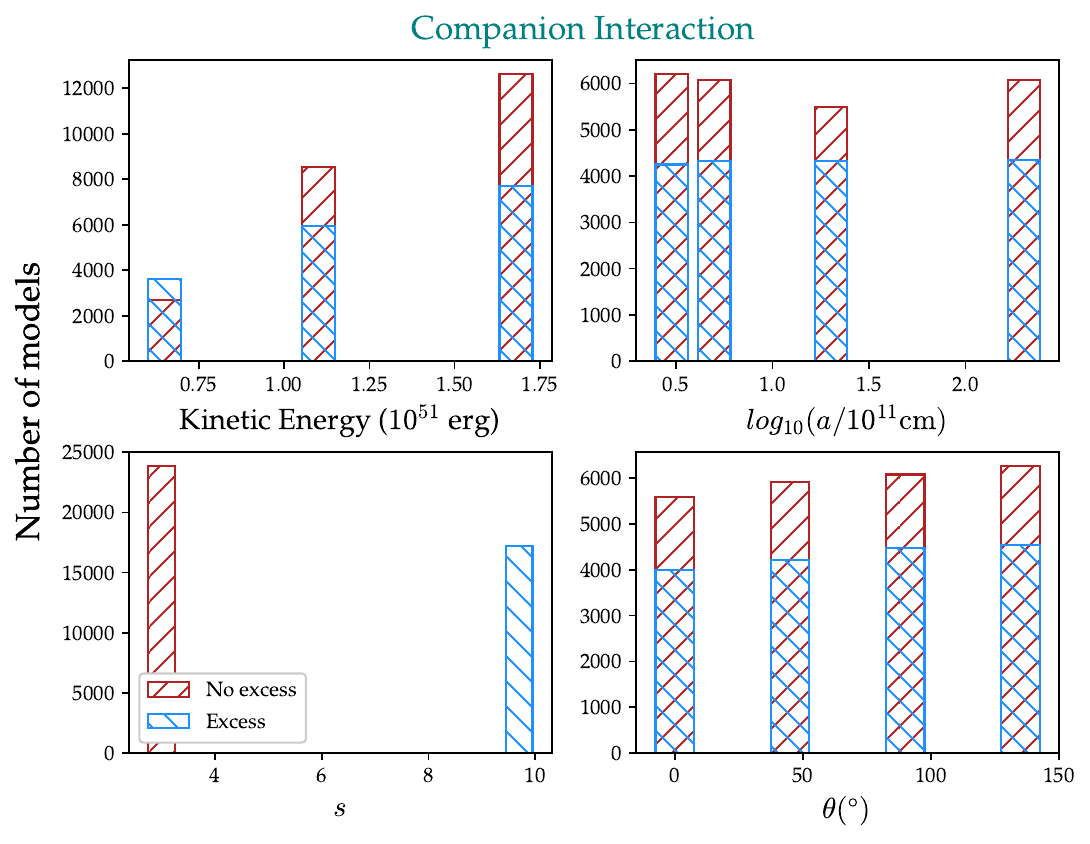}
    \caption{Similar to Fig.~\ref{fig:parameter_distributions_CompanionInteraction} but for non-\eexsne.}
    \label{fig:parameter_distributions_nonEExSNe_CompanionInteraction}
\end{figure}

The parameter distributions for the best-fit DD and CI models for non-\eexsne from Sect.~\ref{subsec:models_with_excess} are presented in Fig.~\ref{fig:parameter_distributions_nonEExSNe_DoubleDetonation} and \ref{fig:parameter_distributions_nonEExSNe_CompanionInteraction}.

\end{appendix}
\end{document}